\documentclass[10pt,a4paper]{article}

\usepackage[utf8]{inputenc}  
\usepackage[T1]{fontenc}   
\usepackage[round]{natbib}
\usepackage{fancyhdr}
\usepackage{dsfont}
\usepackage{indentfirst}
\usepackage{booktabs}
\usepackage{graphicx}
\usepackage{subcaption}
\usepackage{authblk}
\usepackage{parskip}
\usepackage{mathrsfs}
\usepackage{amsmath,amsfonts,amssymb,amsthm}
\usepackage{xcolor}
\usepackage{algorithm}
\usepackage{url}
\usepackage{xparse,nameref}
\usepackage{caption}
\usepackage[flushleft]{threeparttable}
\usepackage{multirow}
\usepackage{tcolorbox}
\usepackage{enumitem}
\usepackage{siunitx}

\theoremstyle{definition}

\newcommand{\ind}{\perp\!\!\!\!\perp}

\usepackage[bottom=2.5cm,top=2.5cm,left=2.5cm,right=2.5cm]{geometry}

\usepackage[utf8]{inputenc}  
\usepackage[T1]{fontenc}

\begin{document}

\providecommand{\keywords}[1]{\textbf{\textit{Keywords --}} #1}

\title{Interpretable ML for High-Frequency Execution}

\author{Timothee FABRE$^{\ast}$$\dag$\thanks{timothee.fabre@centralesupelec.fr} and Vincent RAGEL$^\ast$\thanks{vincent.ragel@centralesupelec.fr}\\
\affil{$\ast$Laboratoire Mathématiques et Informatique pour la Complexité et les Systèmes, CentraleSupélec, Université Paris-Saclay, France\\
$\dag$SUN ZU Lab\\
}}

\maketitle

\begin{abstract}

Order placement tactics play a crucial role in high-frequency trading algorithms and their design is based on understanding the dynamics of the order book. Using high quality high-frequency data and a set of microstructural features, we exhibit strong state dependence properties of the fill probability function. We train a neural network to infer the fill probability function for a fixed horizon. Since we aim at providing a high-frequency execution framework, we use a simple architecture. A weighting method is applied to the loss function such that the model learns from censored data. By comparing numerical results obtained on both digital asset centralized exchanges (CEXs) and stock markets, we are able to analyze dissimilarities between feature importances of the fill probability of small tick crypto pairs and Euronext equities. The practical use of this model is illustrated with a fixed time horizon execution problem in which both the decision to post a limit order or to immediately execute and the optimal distance of placement are characterized. We discuss the importance of accurately estimating the clean-up cost that occurs in the case of a non-execution and we show it can be well approximated by a smooth function of market features. We finally assess the performance of our model with a backtesting approach that avoids the insertion of hypothetical orders and makes possible to test the order placement algorithm with orders that realistically impact the price formation process.
\end{abstract}

\noindent \keywords{Optimal Execution; Fill Probability; Survival Analysis; Limit Order Book; High Frequency}

\setcounter{tocdepth}{2}
\tableofcontents
\newpage

\section{Introduction}

Estimating one's execution probability when providing liquidity is a must either for market making or for optimal execution. Recent promising methods leverage complex deep learning architectures applied to raw data or to synthetic data when data is missing or censored. Yet, using complex features and simple neural architectures is both faster and more interpretable.

In continuous double-auction markets, one can choose to add liquidity by posting a limit order, which carries the risk of potentially not being executed if the market price does not reach the limit price, or to take liquidity by sending a marketable order, which is more expensive due to the crossing of the spread. Depending on the agent's objective and utility function, a trading algorithm may favor one order type over the other and adjust the price of the limit order according to some predefined rules. For example, a market making strategy manages the inventory risk and avoids adverse selection by adjusting the posting distances around a fair price. An execution algorithm seeks for a discounted execution price on one side of the book while aiming at a target quantity to execute under a specific time horizon that can be arbitrarily small, \textit{i.e.} of the order of milliseconds to seconds. Although these strategies are of different nature, they share a central source of uncertainty that is the randomness of execution.

Ultimately, deciding between sending a limit or a market order requires the knowledge of the execution probability of a limit order. The latter depends on market variables such as the bid-ask spread, the volatility, the order flow regime, etc. The order flow itself is highly sensitive to many market variables and they may be clustered in two classes of features: \textit{snapshot variables} which compose the Markovian part of the model (bid-ask spread for example) and \textit{differential variables} which compose the non-Markovian part of the model (traded volume variation computed over a period or realized volatility for example). A fill probability model should find its predictive power in both classes of variables in order to capture the main explanatory features of the order flow dynamics.

When opting for the placement of a limit order in the book, the non-execution risk combined with price volatility may cause additional costs ---the clean-up cost--- if the target quantity to trade is not reached at the end of the time horizon. Assuming that the target quantity must be traded at all costs, a marketable order sent at the end of the period will potentially trade at a worse price than if it had been sent at the beginning due to adverse price movements. The estimation of such a price move in the case of non-execution is capital since it draws the decision boundary of the order placement tactic, and its sensitivity to market variables leads to a complex interaction with the fill probability function in the decision-making process. As an example, one easily intuits that both fill probability and clean-up cost functions increase with the realized volatility which will force the agent to make a trade-off between market risk and execution likelihood. Therefore, to decide between a limit order and a market order, one must properly evaluate both the execution probability and the clean-up cost.

\subsection{Literature review}

The fill probability function estimation is generally carried out using two main classes of methods. The first class encompasses survival analysis tools and is applied to financial data in \citet{cho2000probability, lo2002econometric}, that carry out empirical analysis of the role played by multiple key market variables. The second class of estimation procedure encompasses the insertion of hypothetical limit orders and the computation of their first passage time, either using transaction data or the crossing of a reference price. First passage time distributions of non-Gaussian dynamics are widely used in the literature and their empirical scaling properties are carefully studied in \cite{perello2011scaling}. The tail exponents of the first passage time and empirical time-to-fill and time-to-cancel are analyzed in \cite{eisler2012price} and the authors suggest that the fatter tails observed for the first passage times are explained by cancellations. They also provide a simple model that succeeds in capturing the above stylized facts. A state dependent model of the fill probability function was proposed in \cite{maglaras2022deep} for which a recurrent neural network was used to compute the fill probability of synthetic limit orders as a function of market features. In a more recent work \citep{arroyo2024deep}, the authors develop a deep neural network structure for the estimation of the full survival function. They use the right-censored likelihood as a loss function for training and proper scoring rules for model performance assessment. Although they apply the model to market data, the only interpretable features they use are the spread, the volatility and the best queue imbalance. They provide a feature importance analysis, but the small number of predictions used for the computation of the Shapley values as well as the use of raw volumes and prices as inputs leads to a lack of interpretability in the results.

In \cite{maglaras2022deep}, the authors point out the fact that using real limit orders for fill probability computation brings a selection bias to the analysis, which finds its nature in the heterogeneity of information contained in the order flow. Since limit orders are posted by both informed and uninformed traders with various strategies and time scales, the analysis of such orders cannot be used for the purpose of fill probability computation of an uninformed agent with a specific horizon. The problem is that in practice, the insertion of a limit order impacts the order book and consequently the price process itself, \textit{e.g.} see \citet{weber2005order, hautsch2012market, bacry2016estimation, said2017market, brogaard2019price}. Last but not least, the posted size affects market depth and order flow imbalances and thus the execution probability itself. This is even more true in high frequency trading, and the classical first passage time method fails at capturing these stylized facts as the main hypothesis is the absence of impact of posted limit orders. Following the findings of \cite{lo2002econometric} and the above reasoning, we decide to favor market impact adjusted fill probabilities, \textit{i.e.} real order flow fill probabilities over the removal of selection bias.

Optimal limit order placement tactics are studied in various frameworks \citep{wald2005optimal, avellaneda2008high, laruelle2013optimal, bayraktar2014liquidation, markov2014design, cartea2015optimal, cont2017optimal}. The LOB dynamics are often characterized with an execution intensity that is a decreasing function of the distance. An univariate representation of the execution probability is certainly unrealistic and restrictive, but it still provides enough information to develop order placement algorithms. In \cite{lehalle2017limit}, the authors build a stochastic control framework and study the impact of adverse selection risk and latency on optimal order placement algorithms. Work has also been done to tackle the queue position valuation problem, see \citet{moallemi2016model, donnelly2018optimal}.

In the field of machine learning, recent works have proposed state-of-the-art architectures to tackle survival analysis in high dimension. \cite{lee2018deephit} have proposed a deep neural network for competing risks frameworks and \cite{katzman2018deepsurv} have introduced a more sophisticated approach that integrates the classical Cox model framework with the power of deep learning. More recently \cite{wang2022survtrace} have proposed a model using transformers designed to handle competing events. We decide to opt for simple feed-forward network architecture instead, as we aim at proposing execution frameworks that are suitable to high-frequency trading. Even though complex architectures can perform better, they are generally not adapted to live HF trading environments where an extra microsecond of tick-to-trade latency due to model computations can lead to significant losses.

\subsection{Contribution and organization of the paper}

Instead of using sophisticated deep learning models in order to extract relevant features but may be difficult to train, our approach uses a combination of handcrafted features and  simple feed-forward networks to  capture key dependencies while being easily integrated in a decision-making process for execution. This choice simplifies the model structure, leading to improved transparency and better reproducibility of the results.

Our contributions are as follows:
\begin{itemize}
    \item We define three new microstructure features for fill probability prediction: the limit order flow imbalance, the aggressiveness index and the priority volume. Using survival analysis methods, we provide empirical evidence of a smooth dependence of the fill probability with respect to these variables.
    \item We apply the inverse-probability-of-censoring weighting method (IPCW) to the training of a neural network, enabling it to learn from censored data.
    \item A detailed feature importance analysis is provided, demonstrating key differences between small tick cryptocurrencies and Euronext equities regarding the predictive power of interpretable variables. Moreover, to the best of our knowledge, our work is the first to analyze the feature importance of the fill probability depending on the placement of the order, \textit{i.e.} if the order is placed inside the spread, at the current best queue, or deep in the book.
    \item We set an optimal execution problem where the agent needs to choose between posting a limit order and sending a marketable order. We propose a new backtest methodology that, by essence, takes into account the market impact of limit order insertion for better performance assessment.
\end{itemize}

The structure of this work is as follows: Section 3 presents the high-frequency data sets that will be used in the numerical experiments. Section 4 discusses two survival analysis methods to compute fill probability on level 3 data and provides empirical evidence of smooth non-linearity of the fill probability as a function of market features. Once these ingredients are gathered, we present our fill probability model in Section 5 and finally study an order placement algorithm that incorporates both the fill probability function and a clean up cost model. Results about the optimal placement policy and an execution-specific backtesting approach are displayed before discussing the importance of latency and how it can be integrated in the cost function.

\section{Data}

Our work uses two high-frequency data sets with Level-3 granularity (full order details) for digital assets and a mature equity market.

The digital asset data is provided by SUN ZU Lab's proprietary feed handlers. There are few centralized exchanges who provide a market by order data API amongst which Coinbase, Bitstamp and Bitfinex are the most popular. We chose to use Coinbase data for two reasons. First, the tick size remains very small during the covered period of time which is not the case for Bitstamp for example where tick sizes have been enlarged in 2022. Secondly, Coinbase provides a timestamp with microsecond precision and a sequence number with each message allowing us to be confident about the order book reconstruction process. We have 1 month of data at our disposal, from 2022-11-05 to 2022-12-05 on  BTC-USD and ETH-USD. We removed several days (2022-11-07 to 2022-11-10) from our analysis because of the extreme volatility that was observed during this period. The equity data comes from BEDOFIH (Base Européenne de Données Financières à Haute-fréquence), built by the European Financial Data Institute (EUROFIDAI). This comprehensive data base offers highly detailed order data for all stocks traded on Euronext Paris between 2013 and 2017.  To conduct our analysis, we will concentrate on the most recent year available, from January 2017 to December 2017, for two liquid French stocks, BNPP and LVMH. These  are Level 3 data feeds which contain the full sequence of order-based events. This makes it possible to trace the life of each order and identify when it was completed or canceled. However, it requires to reconstruct the order book from individual order events. Table \ref{tab:descriptive_statistics_general} summarizes several descriptive statistics of the data. The spread is clearly different between small tick assets (BTC-USD, ETH-USD) and large tick assets (BNPP, LVMH). The tick size is 0.01 USD for BTC-USD and ETH-USD pairs representing roughly $10^{-6}$ of BTC-USD and $10^{-5}$ of ETH-USD. The tick size depends on the price value for both BNPP and LVMH with respect to MIFID rules. In \cite{huang2016predict}, \cite{laruelle2019assessing}, the differentiation is made for an average spread limit of 1.6 ticks but we still classify BNPP as a large tick asset with respect to digital assets.

We display some descriptive statistics of the lifetime of orders in Table \ref{tab:descriptive_statistics_lifetime}, expressed in number of seconds. We observe that most of the orders of the cryptocurrency pairs have significantly smaller lifetimes than those of the equities. The discrepancy between the median and the average indicates the presence of fat tailed distributions.

\begin{table}
    \sisetup{group-digits=false}
    \caption{\textit{Descriptive statistics} --- Spread, trade size and daily volume statistics}
    \begin{center}
        \begin{tabular}{clccc}
            \toprule
            \toprule
                    &            & \textbf{Spread}$^{\rm a}$  & \textbf{Trade size}$^{\rm b}$ & \textbf{Daily volume}$^{\rm c}$\\
            \midrule
                     & 5\%        & 84.31   &  1.01 &  197,146,101   \\
                     & Median     & 163.40   &  127.86 &  470,617,696  \\
            BTC-USD  & 95\%       & 394.25  &  8,619.66 &  1,635,066,155  \\
                     & Mean       & 192.51  &  1,725.54 &  612,757,871  \\
            \midrule
                     & 5\%        & 8.31    &  0.84 &  221,517,905 \\
                     & Median     & 15.61    &  371.70 &  431,407,671  \\
            ETH-USD  & 95\%       & 37.34   &  7,915.72 &  1,459,323,150  \\
                     & Mean       & 18.07   &  1,836.98 &  563,289,467 \\
        
            \midrule
                         & 5\%     &         1.00    &        20   &    496,135  \\
                         & Median  &         2.00    &       140   &   194,8473  \\
            BNPP          & 95\%    &         3.00    &       386   &   3,716,829  \\
                         & Mean    &         1.70    &       180   &   2,092,348  \\
            \midrule
                         & 5\%     &         1.00   &         7   &     99,870  \\
                         & Median  &         1.00   &        49   &    319,678  \\
            LVMH         & 95\%    &         2.00    &       147   &    573,315  \\
                         & Mean    &         1.24    &        61   &    324,954  \\
            \bottomrule
        \end{tabular}
    \end{center}
    \begin{tablenotes}
      \small
      \item $^{\rm a}$ Bid-ask spread is expressed in number of ticks.
      \item $^{\rm b}$ Trade size is expressed USD for crypto pairs and in number of stocks for the equities. The metrics are computed on the volume of the recorded market orders and serves as a reference for the average trade size (ATS).
      \item $^{\rm c}$ Daily volume is expressed in USD for crypto pairs and in number of stocks for the equities.
    \end{tablenotes}
    \label{tab:descriptive_statistics_general}
\end{table}

\begin{table}
\sisetup{group-digits=false}
    \caption{\textit{Descriptive statistics} --- Time-to-event statistics, expressed in seconds}
    \begin{center}
        \begin{tabular}{clccc}
            \toprule
            \toprule
                     &                     & \textbf{Lifetime}  & \textbf{Time to fill} & \textbf{Time to cancel}\\
            \midrule
                     & 5\%        & 0.003       & 0.0005    & 0.003 \\
                     & Median     & 0.063      & 0.125    & 0.059 \\
            BTC-USD  & 95\%       & 8.499      & 5.697   & 7.107 \\
                     & Mean       & 152.931      & 53.672   & 4.176 \\
        
            \midrule
                     & 5\%        & 0.003        & 0.0006    & 0.003 \\
                     & Median     & 0.069       & 0.205    & 0.065 \\
            ETH-USD  & 95\%       & 9.035       & 11.274   & 7.839 \\
                     & Mean       & 117.453       & 65.914   & 4.347 \\
        
            \midrule
                         & 5\%     &       0.0004   &       0.0007   &       0.0004   \\
                         & Median  &        1.296   &        4.069   &        1.193   \\
            BNPP         & 95\%    &       100.08   &       106.56   &        96.31   \\
                         & Mean    &        48.19   &        70.73   &        32.68   \\
        
            \midrule
                         & 5\%     &       0.0002   &       0.0009   &       0.0002   \\
                         & Median  &        4.173   &        8.375   &        3.983   \\
            LVMH         & 95\%    &       292.92   &       231.69   &       281.28   \\
                         & Mean    &        85.13   &        84.91   &        70.16   \\
        
            \bottomrule
        \end{tabular}
    \end{center}
    \label{tab:descriptive_statistics_lifetime}
\end{table}

\section{Non-parametric analysis of the execution and cancellation risks} \label{sec:non_parametric_analysis}

The role of this section is to introduce three new microstructural features and show that they strongly influence  both executions and cancellations. A non-parametric estimation is carried out using a competing risks framework that we first briefly review.

\subsection{Cancellation as a competing risk}

In the rest of the paper, we place ourselves in a filtered probability space $\left(\Omega,\mathcal{F},\mathbb{F}:=(\mathcal{F}_t)_t,\mathbb{P}\right)$. Let the state of a pending limit order be modeled by the $\mathbb{F}$-adapted process $(X_t)_{t\geq0}$ with values in a discrete state space $\mathbb{S}$ and such that $X_0=0$ $a.s.$. The order is said to be alive at time $t$ if $X_t=0$ and dead if $X_t\neq0$. Hence, the lifetime of this order is fully characterized by the random variable $L:=\text{inf}\{t>0,X_t\neq0\}$. Naturally, all the states (except for 0) are absorbing since they all signify the death of the underlying order. We now discuss the specification of the state space $\mathbb{S}$ in the continuous trading paradigm. We will also denote, for $t\geq0$, $F(t):=\mathbb{P}(L\leq t)$ and $S(t):=1-F(t)$ respectively the cumulative distribution function and the survival function (or complementary cumulative distribution function) of the order lifetime $L$.

This limit order is seen as a birth-death entity for which death ---removal from the order book--- can be triggered either by its full execution or by its cancellation. It is noteworthy that the case of partial execution is not considered here and should be tackled differently since it is not an absorbing state, the order remaining alive in the book. When the death of the order is not observed, we say that the order is right-censored. Censoring is present in both financial markets and crypto CEXs data sets. An example of a censored order is when it is neither cancelled nor fully executed at the end of the observation period. From the practitioner's point of view, the one of the main causes of censoring in CEXs is feed handler disconnections that may happen depending on the venue (this also happens on traditional financial markets but to a lesser extent). In the case of a disconnection, data feed is stopped for a random time that ranges from milliseconds to seconds or even hours for API shutdowns. The censoring issue in the equity dataset occurs at the close of the trading day and is therefore much smaller.

As outlined in \cite{eisler2009diffusive}, the presence of cancellations plays a significant role in the difference observed between empirical first passage time (FPT) and time to fill (TTF) distributions, leading to fatter tails for the FPT. Since we want to analyze both execution and cancellation probabilities, we treat cancellation as a competing risk with respect to execution rather than as censoring since once the order is cancelled, its future execution cannot happen anymore.
In fact, when censoring occurs, the event of interest may still happen afterward, but it goes unobserved.

Based on this, we specify the state space as $\mathbb{S}:=\{0, 1, 2\}$ and define $\mathbb{S}^\dagger:=\{1, 2\}$ the death state space, where 1 indicates a fully executed order and 2 indicates a cancelled order. We define the cause-specific hazard rate for $t\geq0$, $i\in\mathbb{S}^\dagger$
\begin{equation}\label{eq:cause_specific_hazard_rate}
    \lambda_i(t)\,:=\,\lim_{\Delta t \to 0^+} \frac{\mathbb{P}\big(L\in[t,t+\Delta t[, X_L=i|L\geq t\big)}{\Delta t}.
\end{equation}

We define the cumulative incidence function (CIF), for $t\geq0$, $i\in\mathbb{S}^\dagger$
\begin{align}
    F_i(t)&:=\mathbb{P}\left(L\leq t,X_L=i\right)\\
    &=\,\int_0^t\mathbb{P}\left(L>s\right)\lambda_i(s)\mathrm{d}s,
\end{align}
where the last equality is established with elementary calculus using Equation \eqref{eq:cause_specific_hazard_rate}.

\subsection{Non-parametric estimation}

Assume that we either observe the lifetime or the censoring of $N$ limit orders which are exposed to the following mutually exclusive causes of death: execution and cancellation.

Previous notations are naturally indexed in the following manner: we denote the lifetime of the $n$th order by $L_n$ and the distribution function of $L_n$ is denoted by $F$; we assume that the order can be subject to independent right censoring modeled by a random variable $C_n$ and the censoring variable is assumed independent from $L_n$. Thus, what we effectively observe is the realization of the random variables $T_n:=\text{min}(L_n,C_n)$ and $\mathds{1}_{\{L_n\leq C_n\}}$ and $t_n$ denotes our observation of the censored lifetime $T_n$.

Let $\left(t_{(k)}\right)_{1\leq k \leq K}$ be the ordered sequence of observed and censored lifetimes such that for $1\leq n \leq N$, $t_n\in\{t_{(k)}, 1\leq k \leq K\}$. In the seminal paper \cite{kaplan1958nonparametric}, the Kaplan-Meier estimator was introduced as a non-parametric estimator of the survival function in the presence of censored data and is written as
\begin{equation}\label{eq:kaplan_meier_estimator}
    \forall\,t\geq0,\hspace{0.2cm}\widehat{S}(t)=\prod_{k,t_{(k)}<t}\left(1-\frac{d_k}{n_k}\right),
\end{equation}
where $d_k$ is the number of deaths at time $t_{(k)}$ and $n_k$ is the number of pending orders at time $t_{(k)}^-$. Thus, after each $t_{(k)}$, the number of pending orders becomes $n_{k+1}=n_k-(c_k+d_k)$ where $c_k$ is the number of right censored observations occurring at $t_{(k)}$.

A non-parametric estimator of the cumulative incidence function was proposed in \cite{aalen1976nonparametric} and \cite{aalen1978empirical}. The Aalen-Johansen estimator is defined, for $t>0$, and $i\in\mathbb{S}^\dagger$ as

\begin{equation}\label{eq:aj_estimator}
    \widehat{F}_i(t)\,=\,\sum_{k,t_{(k)}<t}\widehat{S}\left(t_{(k-1)}\right)\,\frac{d_k^i}{n_k}
\end{equation}
using the convention $t_{(0)}:=0$ and denoting by $d_k^i$ the number of deaths from cause $i$ observed at $t_{(k)}$. Naturally, the Kaplan-Meier curve $\widehat{S}(.)$ is computed without distinguishing between the causes of death, such that for $1\leq k\leq K$, $d_k=d_k^1+d_k^2$. 
The Aalen-Johansen estimator is built by summing the product of the Kaplan-Meier survival function and increments of the Nelson-Aalen cumulative hazard rate estimator, see \cite{aalen1978nonparametric}, \cite{nelson1969hazard} and \cite{nelson1972theory}.

Counting process theory provides mathematical expressions of confidence intervals for the Aalen-Johansen estimator. The procedure used here is based on the Gray estimator \cite{pintilie2006competing} of the variance of the CIF, written for $t>0$, $i\in\mathbb{S}^\dagger$,

\begin{align}
    \widehat{\text{Var}}\left(\widehat{F}_i(t)\right)\,&=\,\sum_{k,t_{(k)}<t}\frac{\left(\widehat{F}_i(t)-\widehat{F}_i\left(t_{(k)}\right)\right)^2d_k}{(n_k-1)(n_k-d_k)}+\sum_{k,t_{(k)}<t}\frac{\widehat{S}\left(t_{(k-1)}\right)^2d_k^im_k^i}{(n_k-1)n_k}\notag\\
    &\hspace{0.5cm}-2\sum_{k,t_{(k)}<t}\frac{\left(\widehat{F}_i(t)-\widehat{F}_i\left(t_{(k)}\right)\right)\widehat{S}\left(t_{(k-1)}\right)d_k^im_k^i}{(n_k-d_k)(n_k-1)},
\end{align}
where we used the notation $m_k^i:=\displaystyle\frac{n_k-d_k^i}{n_k}$.\\

This estimator is known to over-estimate the true variance \citep{braun2007comparing} slightly. As suggested by \cite{kalbfleisch2011statistical}, we use the log-log method to compute the confidence interval. This methodology restricts the boundaries of the confidence interval to $[0,1]$ as this constraint may break when linear confidence intervals are computed. Let $\alpha$ be the confidence level and $z_x$ the Gaussian $x$-percentile. If we define, for $t>0$ and $i\in\mathbb{S}^\dagger$

\begin{equation}
    C_\alpha^i(t):=z_{\frac{\alpha}{2}}\frac{\sqrt{\widehat{\text{Var}}\left(\widehat{F}_i(t)\right)}}{\widehat{F}_i(t)\log\left(\widehat{F}_i(t)\right)},
\end{equation}
then an $\alpha$-confidence interval $\text{CI}_\alpha(\widehat{F}_i(t))$ is obtained using

\begin{equation}\label{eq:confidence_intervals_aj}
    \text{CI}_\alpha\left(\widehat{F}_i(t)\right)\,:=\,\left[\widehat{F}_i(t)^{e^{-C_\alpha^i(t)}}, \widehat{F}_i(t)^{e^{C_\alpha^i(t)}}\right].
\end{equation}

\subsection{Empirical analysis}

In the following experiments, we analyze both passive limit order flow, \text{i.e.} limit orders that do not affect the bid-ask spread when posted, and aggressive limit order flow, \textit{i.e.} limit orders which form a new best queue when posted. In the latter case, marketable limit orders that are partially executed are not studied and are left for further investigation; note that they could be treated as a special case of aggressive limit orders  since they change both best prices. We decide to discard orders that were posted too far in the book and set the threshold at 10\% on either side of the mid price for cryptocurrencies. As regards equities, we focus on the first 10 limits of the limit order book.

Our goal is to investigate the influence of microstructural features on the fill probability. To this effect, the order fill time horizon must be small, but sufficiently large to be able to characterize the execution risk. Choosing a too short time horizon for the fill probability would lead to extremely imbalanced classes, resulting in high (relative) estimation error. The time horizon depends on the asset considered and especially on the frequency of trades. We observed that a 1 second time horizon for the crypto pairs and a 10 seconds time horizon for the equities gave satisfactory and comparable results. Given these time horizons, the percentage of observed executions amongst posted limit orders is 2\% for the crypto pairs and 4\% for the equities.

Let us now introduce three new variables and analyze the sensitivity of the fill probability and cancellation probability with respect to these variables. We use the Aalen-Johansen estimator defined in Equation \eqref{eq:aj_estimator}, and 95\% confidence intervals are computed using Equation \eqref{eq:confidence_intervals_aj}.

\subsubsection{Limit order flow imbalance}

This measure quantifies the volume imbalance of limit orders that were posted in the last $m$ events, right after the current order insertion, and indicates if the new bid/offer intentions are concentrated on one side of the LOB. Note that it is different from the order flow imbalance used in \cite{su2021price} and \cite{cont2023cross} as cancellations and trades are not used for its computation. At the insertion of the order, we sum the volumes inserted on the bid side $\Delta_m Q_{\text{bid}}$ and on the ask side $\Delta_m Q_{\text{ask}}$ over the last $m$ events (including this order). Notice that $\Delta_m Q_{\text{bid}} \geq 0$, $\Delta_m Q_{\text{ask}} \geq 0$ and $\Delta_m Q_{\text{bid}} + \Delta_m Q_{\text{ask}} > 0$ since the current posted limit order is taken into account. We thus define

\begin{equation}
    \mathcal{I}_{\text{add}}^m\,:=\,\frac{\Delta_m Q_{\text{bid}} - \Delta_m Q_{\text{ask}}}{\Delta_m Q_{\text{bid}} + \Delta_m Q_{\text{ask}}}.
\end{equation}

To our knowledge, our work is the first to study the influence of this variable on fill probability. The number of events $m$ is set to 50. The window size was chosen in order to obtain smooth monotonic probability functions. It would be straightforward to extend this indicator in order to give more weights to both relevant limits and more recent observations. 

\begin{figure}[!ht]
    \centering
    \subfloat[Execution probability]{%
        \includegraphics[width=0.4\linewidth]{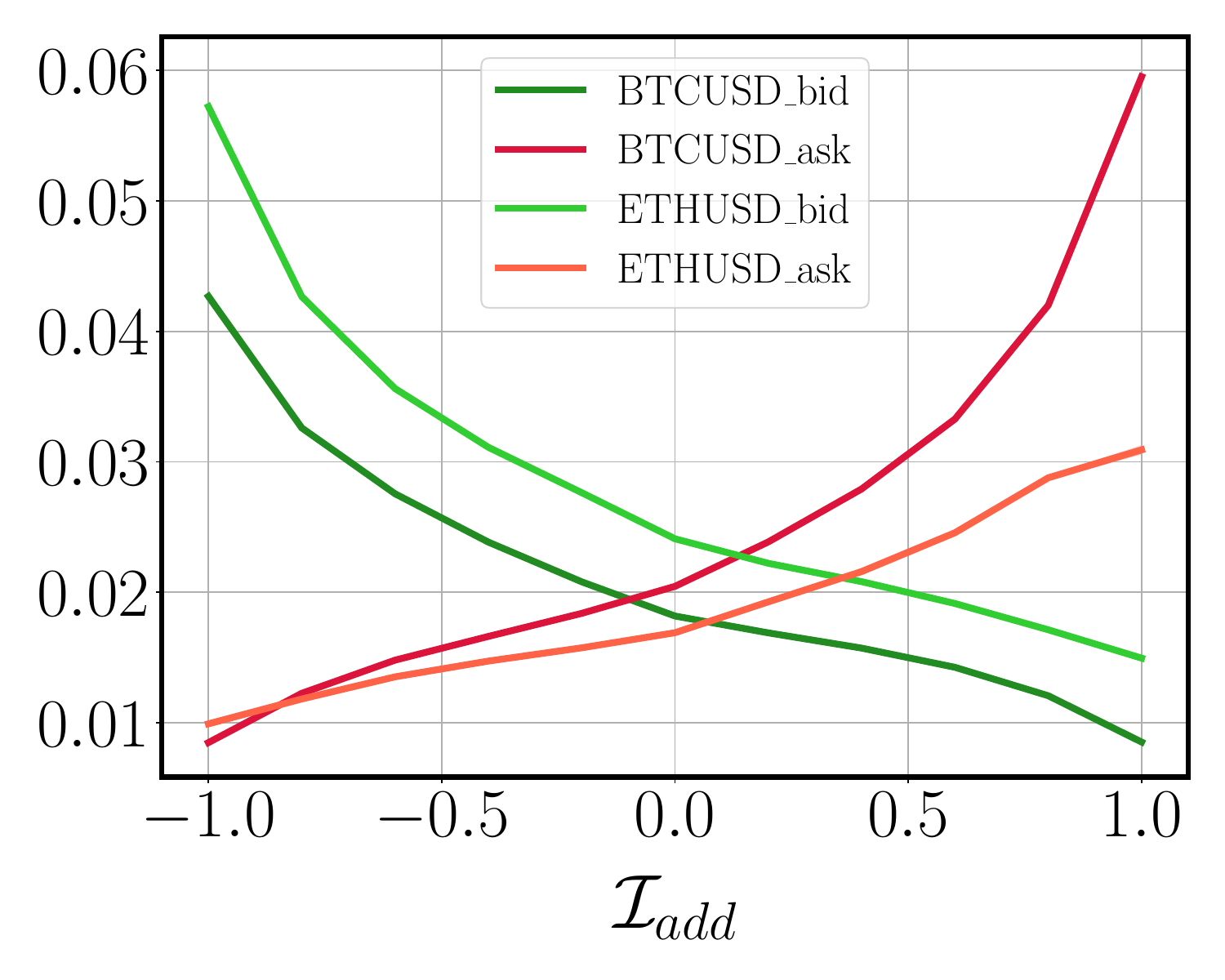}%
    }
    \subfloat[Cancellation probability]{%
        \includegraphics[width=0.4\linewidth]{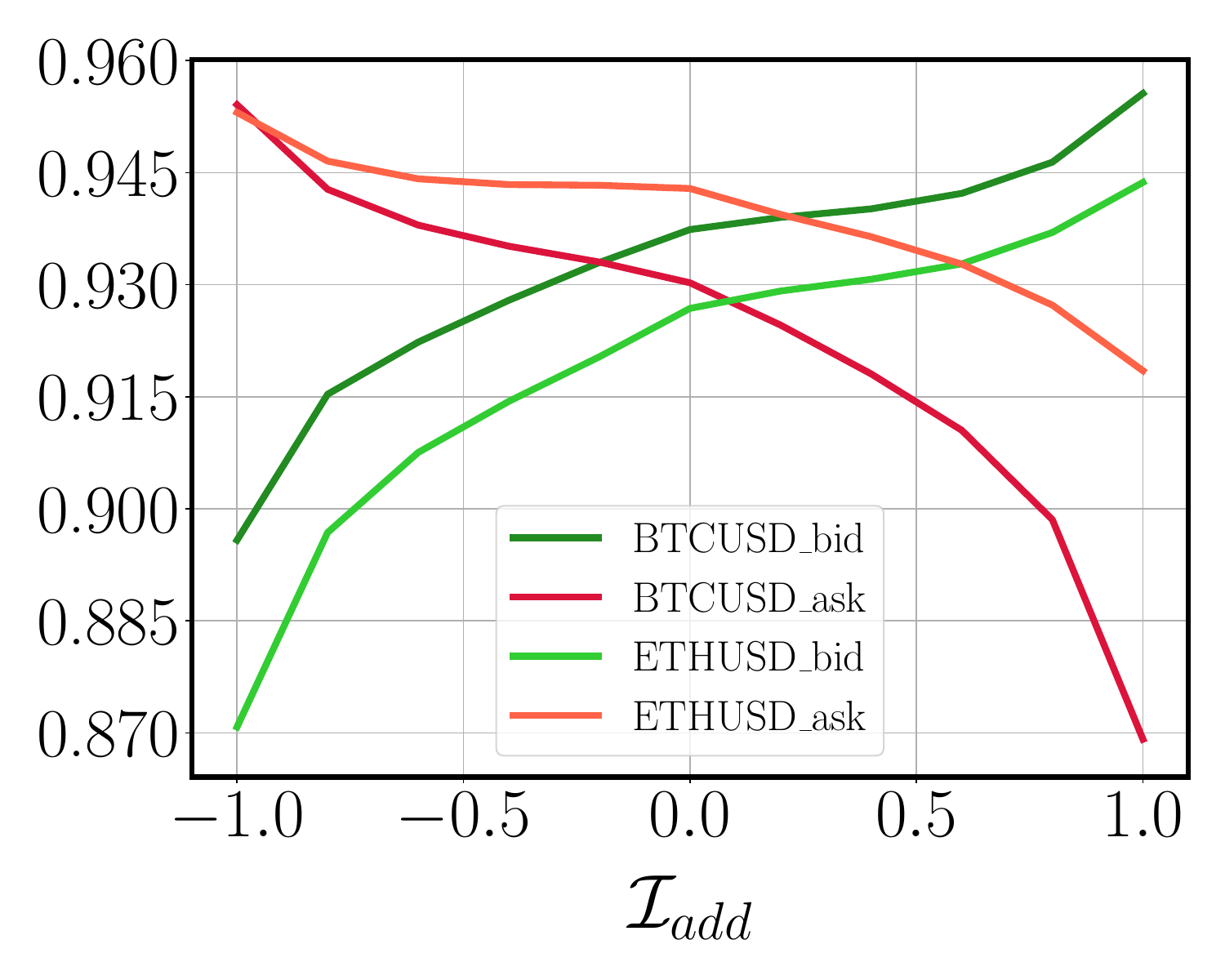}%
    }
    \caption{\textit{Non-parametric analysis} --- 1 second execution and cancellation probabilities as functions of the limit order flow imbalance $\mathcal{I}_{\text{add}}$ measured of the last 50 events.}
    \label{fig:aj_estimator_lof_imbalance}
\end{figure}

The results for BTC-USD and ETH-USD are displayed in Figure \ref{fig:aj_estimator_lof_imbalance}. We observe that the fill probability is symmetrically monotonous as a function of the limit order flow imbalance, which is similar to the shape that one would obtain using the best queue imbalance. For the cancellation probability, we observe an inverse relationship, indicating that agents strongly condition their cancellation policy on measures of market depth variation. Note that we did not manage to obtain a similar empirical evidence for the equities, which suggests that the predictive power of this measure could vary from one asset type to another. 

\subsubsection{Aggressiveness index}

For the purpose of analyzing the fill probability of aggressive order flow, \textit{i.e.} orders that are posted inside the spread, we define a metric that will quantify the degree of aggressiveness of a newly inserted order. We therefore place ourselves in the case of a bid-ask spread before insertion that is (much) greater than 1 tick, limiting the scope of application of the new indicator to small-tick assets. We include the orders that are posted at touch, \textit{i.e.} at the current best queue. In that case, they are passive but can be classified as aggressive orders with a zero aggressiveness index for practicality.

Let $\psi$ be the bid-ask spread before the insertion of the order, and let $\delta$ be the distance of the order with respect to the best queue, such that $\delta>0$ if the order is inserted inside the book, $\delta=0$ if it is posted at touch, and $-\psi<\delta\leq -1$ if it is aggressive. Both $\psi$ and $\delta$ are expressed in number of ticks. We define the aggressiveness index as follows

\begin{equation}
    \omega\,:=\,\frac{\delta}{1 - \psi},
\end{equation}
for $-\psi<\delta\leq-1$.

An index $\omega=0$ corresponds to an order posted at the current best price while a value of 1 indicates a narrowing of the bid-ask spread to its minimal value, \textit{i.e.}, 1 tick. This measure is also expressed in terms of the new bid-ask spread after insertion $\psi^-$, such that $\psi^-\leq\psi$, using the equality

\begin{equation} \label{eq:aggressiveness_index}
    \omega\,=\,\frac{\psi-\psi^-}{\psi-1}.
\end{equation}

\begin{figure}[!ht]
    \centering
    \subfloat[BTC-USD]{%
        \includegraphics[width=0.4\linewidth]{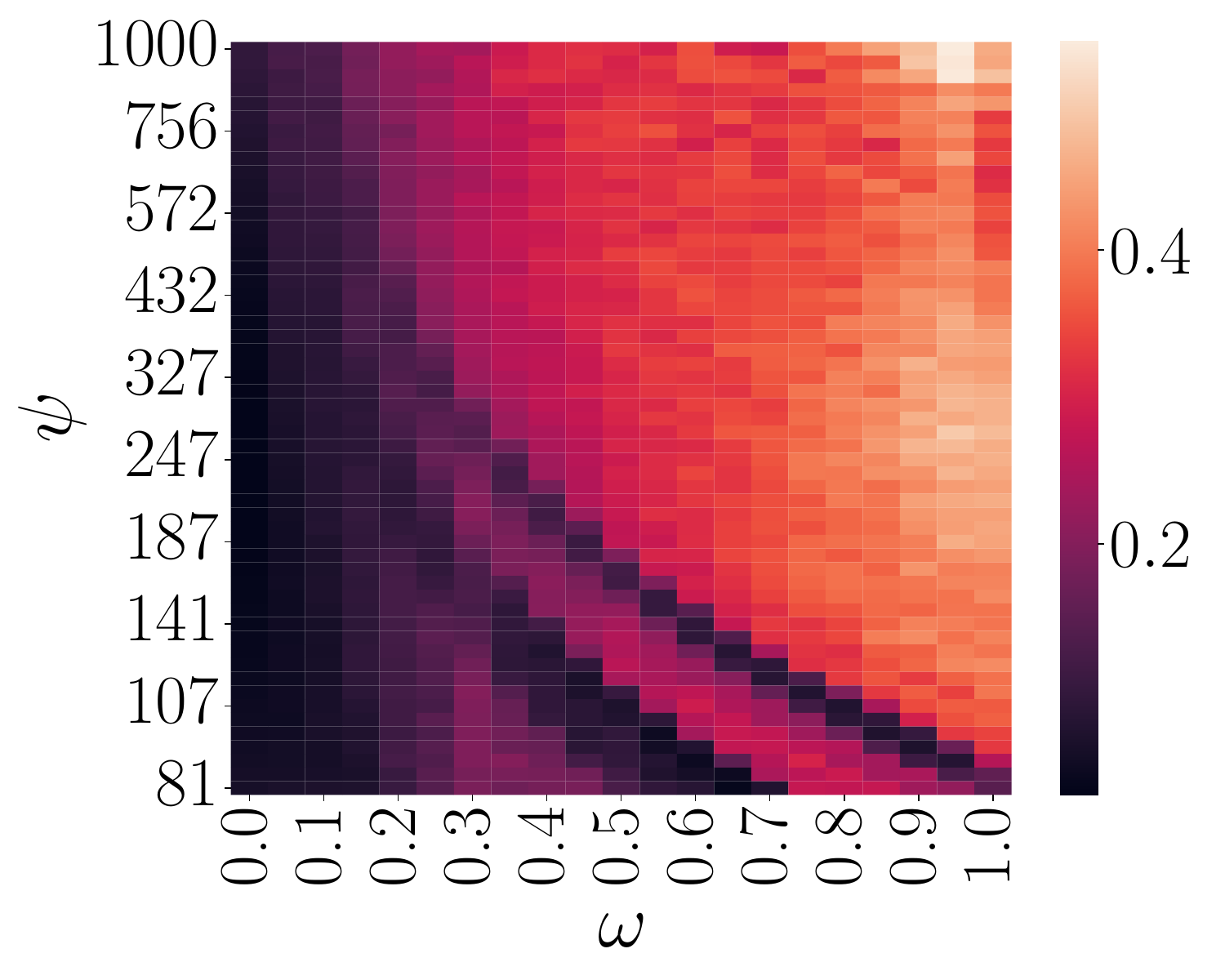}%
    }
    \subfloat[ETH-USD]{%
        \includegraphics[width=0.4\linewidth]{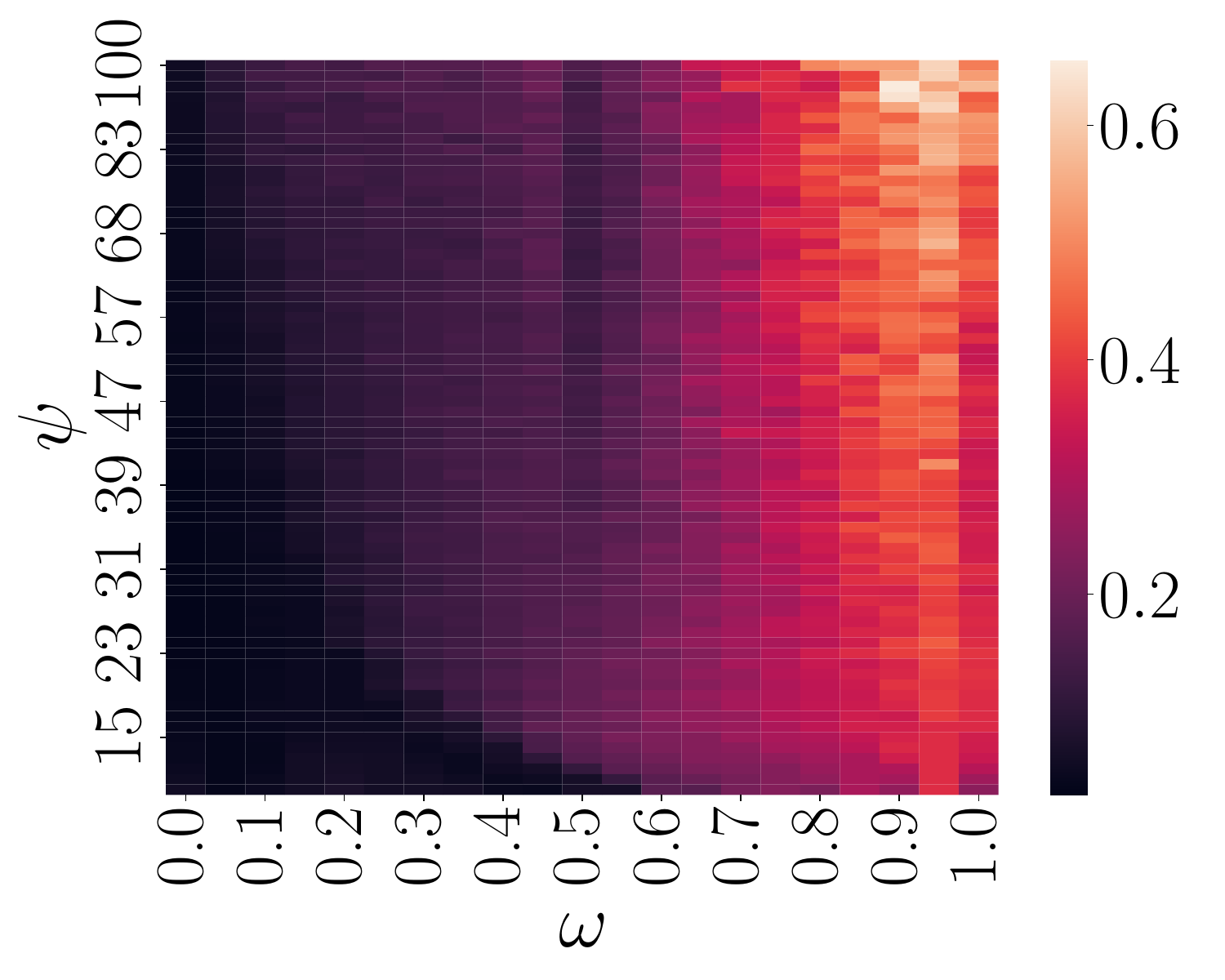}%
    }
    \caption{\textit{Non-parametric analysis} --- 1-second execution probability as a function of the aggressiveness index $\omega$ and the bid-ask spread $\psi$ expressed in number of ticks.}
    \label{fig:aj_estimator_agg_index_exec}
\end{figure}

\begin{figure}[!ht]
    \centering
    \subfloat[BTC-USD]{%
        \includegraphics[width=0.4\linewidth]{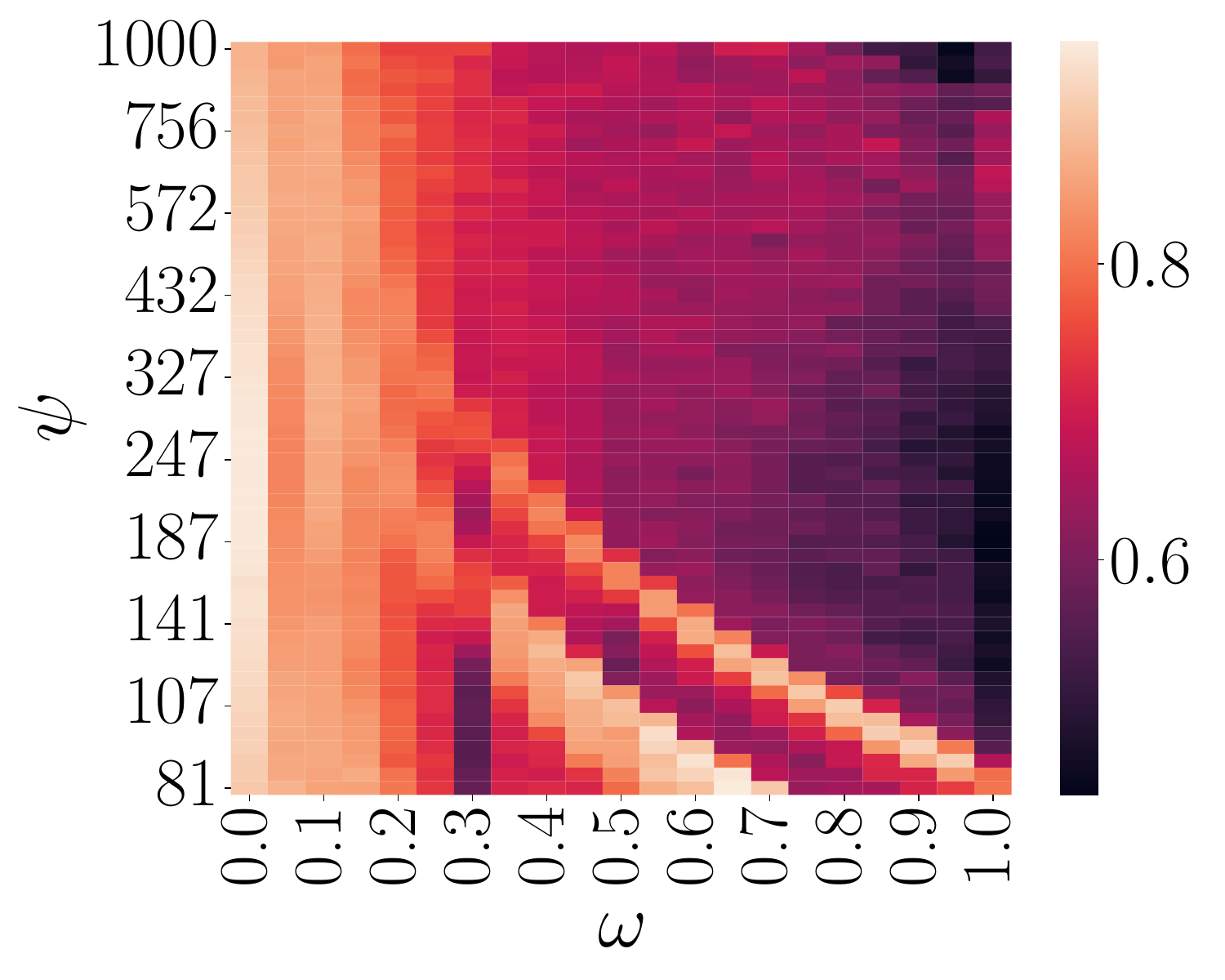}%
    }
    \subfloat[ETH-USD]{%
        \includegraphics[width=0.4\linewidth]{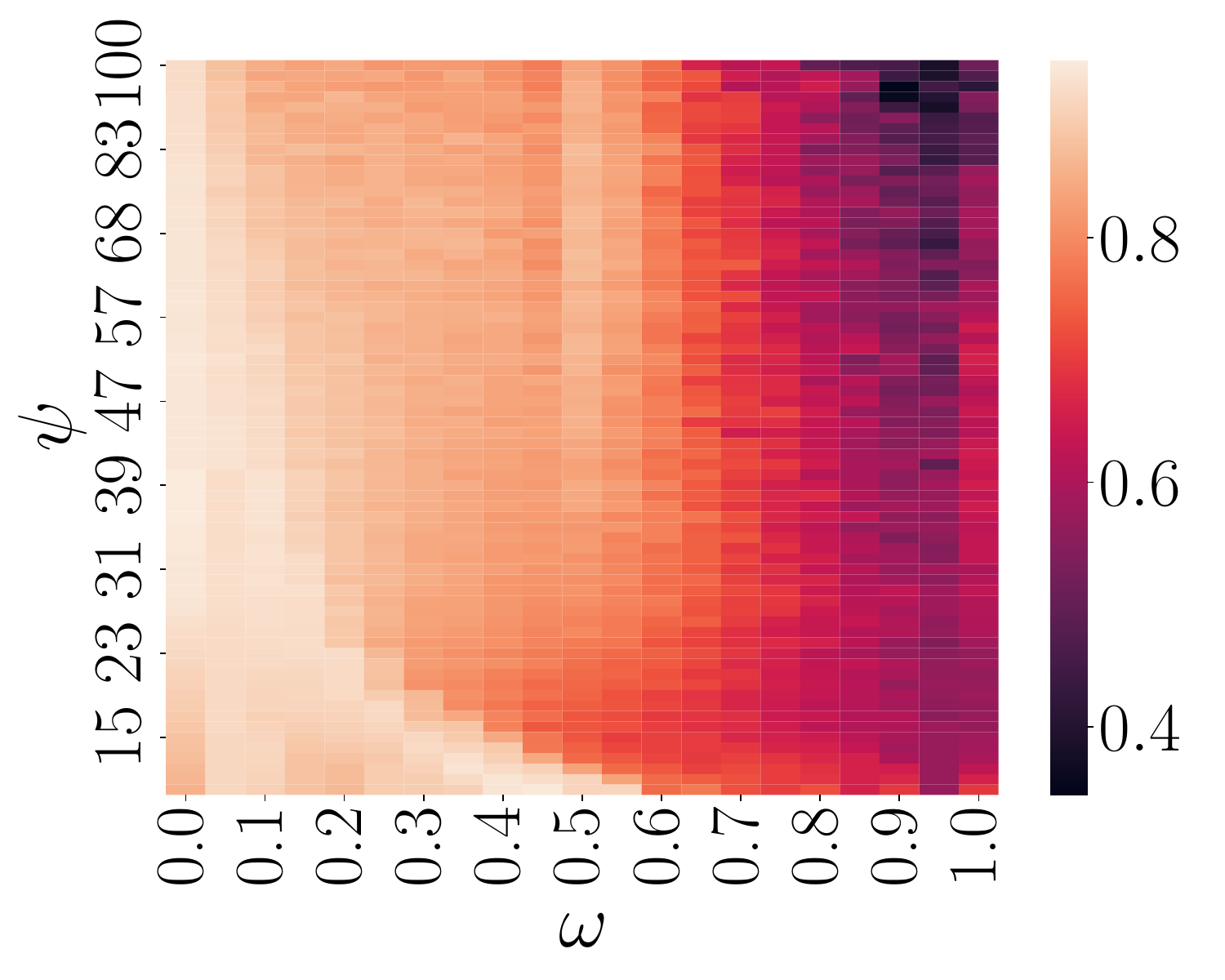}%
    }
    \caption{\textit{Non-parametric analysis} --- 1 second cancellation probability as a function of the aggressiveness index $\omega$ and the bid-ask spread $\psi$ expressed in number of ticks.}
    \label{fig:aj_estimator_agg_index_cancel}
\end{figure}

When posting an order inside the bid-ask spread (and thus creating a new best queue), traders expect a higher execution probability and thus minimize the risk of non-execution while trading at a better price than if they had sent a market order and saving the taker fees. For small-tick assets, it is almost always possible to quote inside the spread and create a new best queue, as the bid-ask spread is generally greater than one tick. Therefore, traders who want to execute fast with minimum slippage may choose to place aggressive orders, leading to a significant tightening of the bid-ask spread as there many of them are competing at the same time. As outlined in \cite{eisler2009diffusive}, aggressive orders will cause an instantaneous rise in the intensity of liquidity taking and will rapidly become like any order resting in the best queue.

The results  for the execution probability are shown in Figure \ref{fig:aj_estimator_agg_index_exec}, and those for the cancellation probability in Figure \ref{fig:aj_estimator_agg_index_cancel}. We observe that the greater the aggressiveness index, the higher the fill probability, and the more aggressive the order is, the less likely it is to be cancelled. This emphasizes at least two phenomena: first, the propensity of impatient agents to optimize their price priority by placing their order at a better price than the current best limit. By doing so, a feedback effect happens: multiple orders are successively inserted in front of each other as the loss of price priority forces aggressive agents to cancel their order and place it again and so forth. The second one is the pinging activity in crypto venues, where orders are submitted inside the spread, at $\delta=-1$, and ca and cancelled shortly thereafter.

\subsubsection{Priority volume}

The priority volume, denoted by $V_{prior}$ is computed by summing the volume of orders at better prices and those at the same price level with better time priority. This metrics complements the distance of placement of the limit order, denoted by $\delta$. The smaller the priority volume, the greater the priority of execution when a marketable order hits the book.

\begin{figure}[!ht]
    \centering
    \subfloat[Execution probability]{%
        \includegraphics[width=0.4\linewidth]{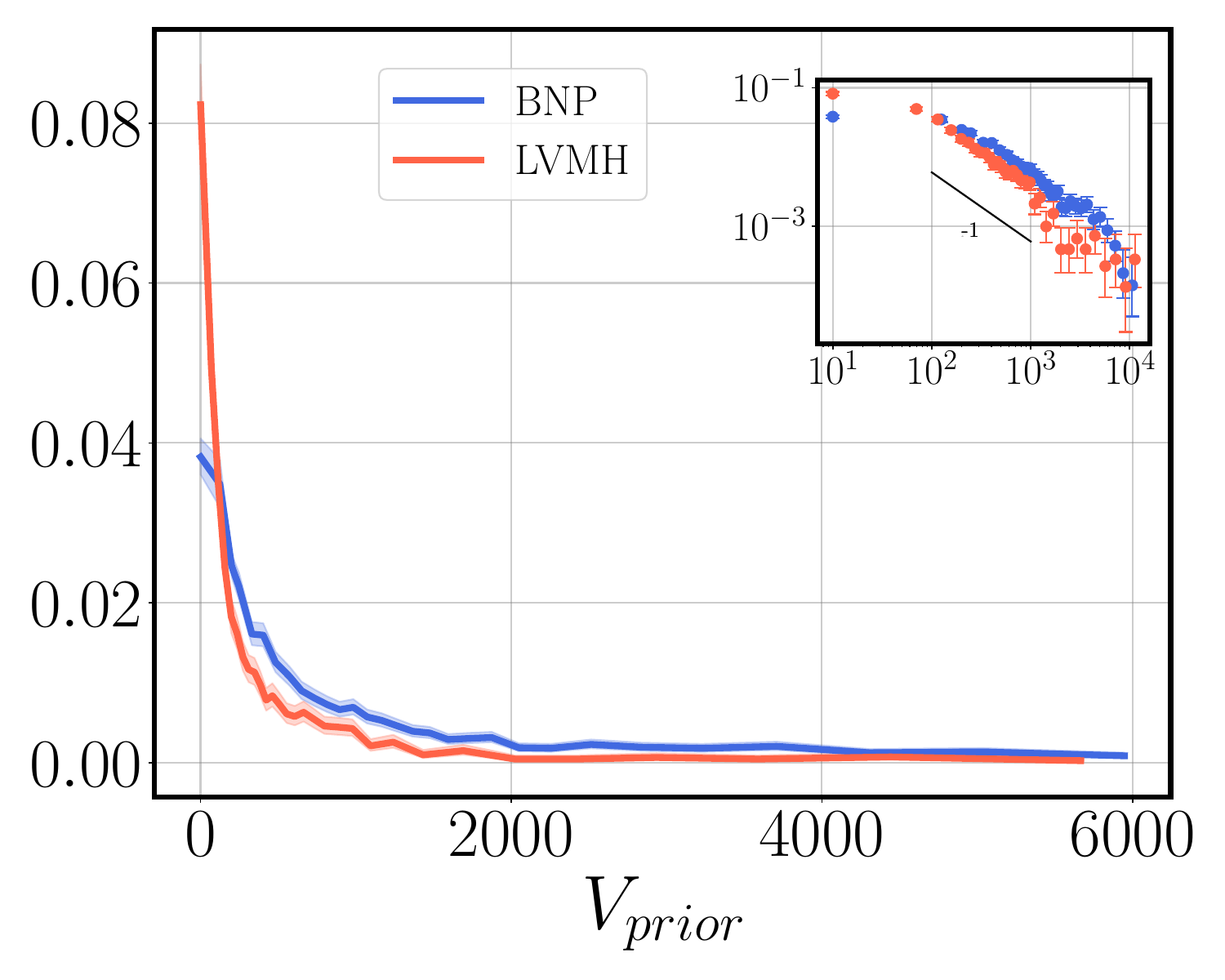}%
    }
    \subfloat[Cancellation probability]{%
        \includegraphics[width=0.4\linewidth]{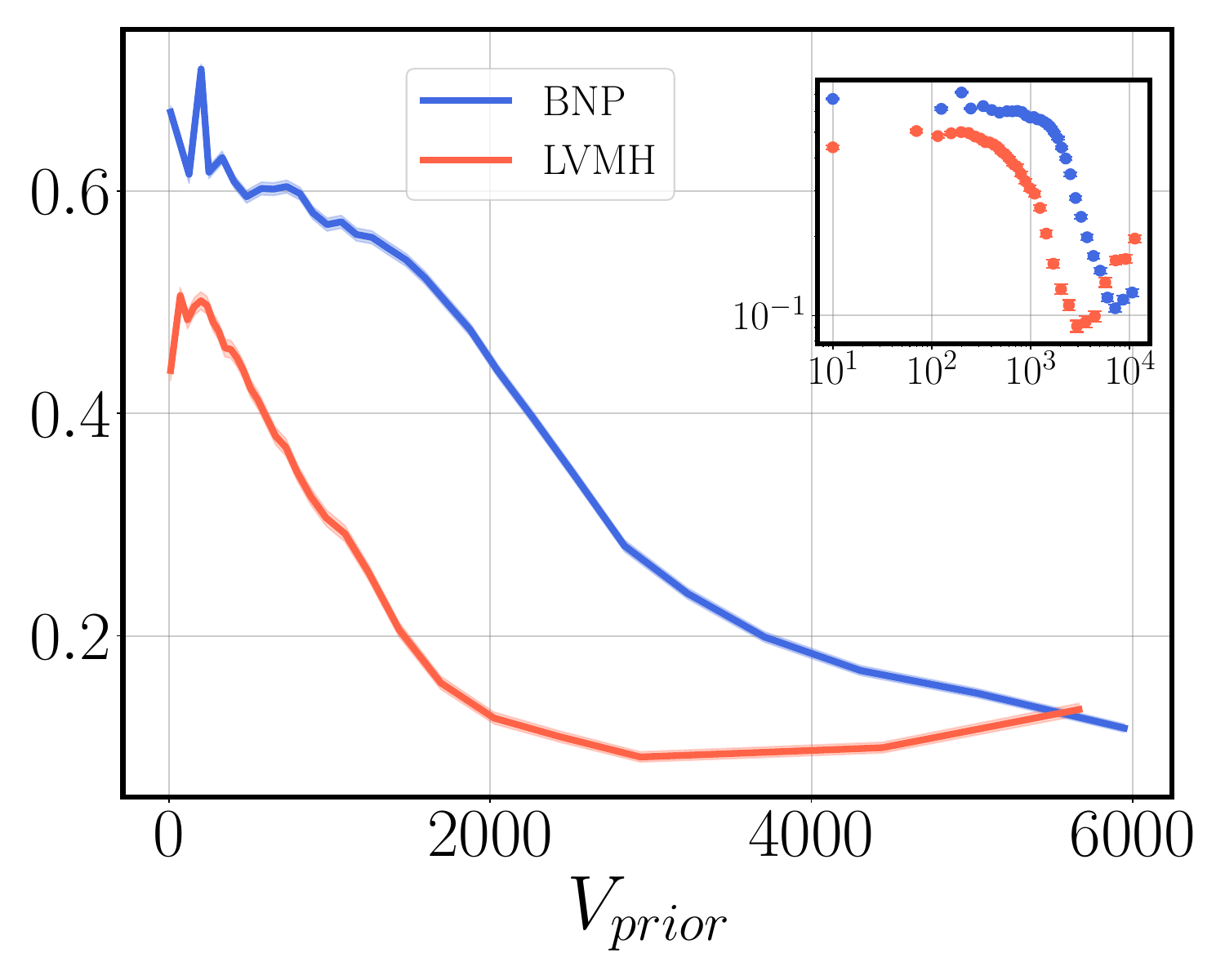}%
    }
    \caption{\textit{Non-parametric analysis} --- 10-second execution and cancellation probabilities as functions of the priority volume $V_{\text{prior}}$ of the order. The volume is expressed in number of shares.}
    \label{fig:aj_estimator_prior_volume_equity}
\end{figure}

\begin{figure}[!ht]
    \centering
    \subfloat[Execution probability]{%
        \includegraphics[width=0.4\linewidth]{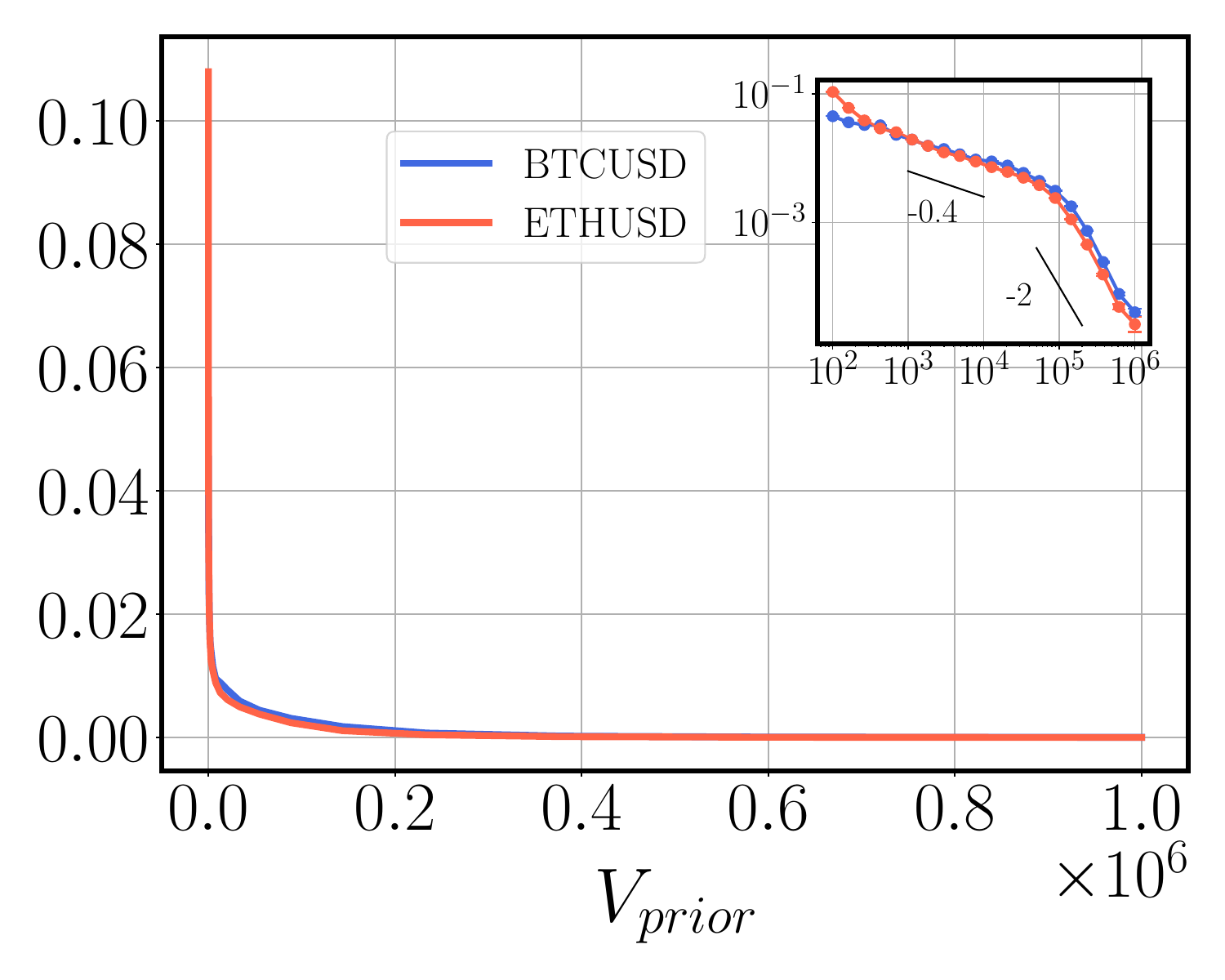}%
    }
    \subfloat[Cancellation probability]{%
        \includegraphics[width=0.4\linewidth]{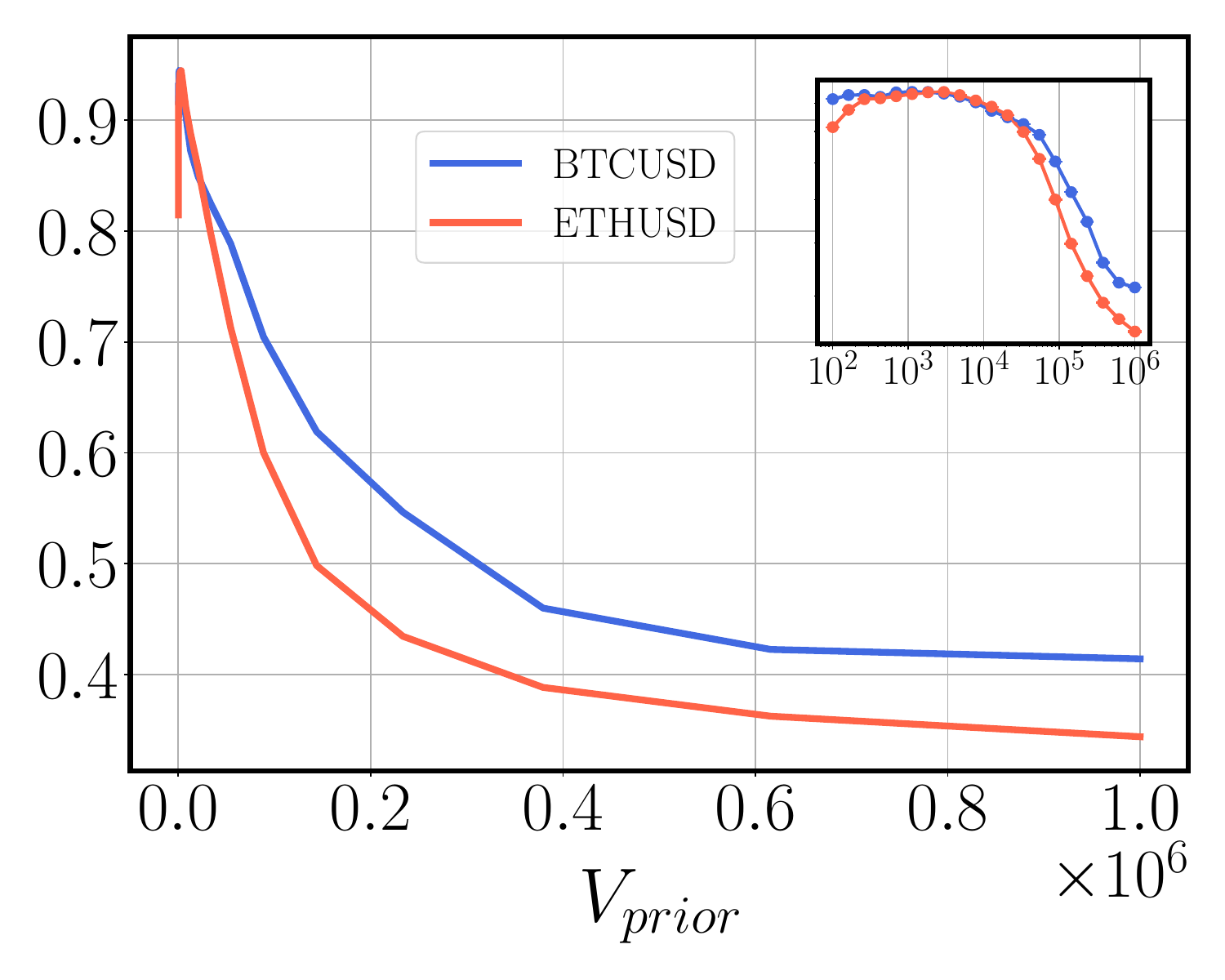}%
    }
    \caption{\textit{Non-parametric analysis} --- 1-second execution and cancellation probabilities as functions of the priority volume $V_{\text{prior}}$ of the order. The volume is expressed in USD.}
    \label{fig:aj_estimator_prior_volume_crypto}
\end{figure}

Results are displayed in Figures \ref{fig:aj_estimator_prior_volume_equity} for the equities and \ref{fig:aj_estimator_prior_volume_crypto} for the crypto pairs. Functions proportional to  $V_{prior}^{-\alpha}$ with $\alpha\in\{0.4, 1, 2\}$ that appear as straight lines in a log-log plot are added for visual reference. Remarkably, the fill probability functions of the two equities are very similar, and those of the crypto pairs are too. For both asset classes, the probability decreases slowly with respect to the volume, but further investigations are needed to validate a specific parametric form. It is noteworthy that there is a major difference between both asset classes concerning the mechanisms that lie behind the execution of orders with respect to the prior pending liquidity. Indeed, small-tick limit order books are sparse, meaning that there are many gaps of liquidity within them; in other words, many prices are not quoted in such LOBs, whereas the price limits of large tick assets are generally quoted up to some market depth. Thus, while the distance is a reasonable proxy for the price priority of an order in a large tick book, it can be misleading for small tick books as a large distance could be coupled with a small priority volume.

\section{A fill probability model for tactical order placement} \label{sec:fill_probability_model}

Our aim is to train an simple artificial neural network with well-chosen features and using raw data only. This makes it possible to use explainable AI to interpret the influence of each feature. We then apply it to optimal order placement and propose a backtest methodology to assess the performance of this type of strategy.

\subsection{A note on the data used for training fill probability models}

Our goal is to estimate  from a feature vector $Z$ the fill probability over a fixed time horizon $T$ that we set to 1 second for crypto pairs and 10 seconds for equities. Our main assumption is that the order is not cancelled within the time horizon $T$. Nevertheless, instead of simply discarding the limit orders that are cancelled within the time horizon, we keep them for the loss weighting procedure that will be described later. Our procedure differs from \cite{maglaras2022deep} in that we do not generate any synthetic order but rather train the model on historical order flow. Despite our exposition to selection bias, the benefit of this approach is threefold.

Firstly, as pointed out in \cite{lo2002econometric}, fill probabilities of hypothetical limit orders do not lead to accurate estimates of the actual fill probability. While this is quite clear for market orders, posting a limit order also causes market response and price impact (see e.g. \cite{eisler2012price} for a consistent price impact analysis). Using a first passage time method would certainly inflate the true fill probability. Things get even worse when posting the order near the mid price since it modifies the liquidity imbalance, a key feature in next trade sign prediction. If computed on “infinitesimal” orders, using such a fill probability model with orders that may inverse the best liquidity imbalance would lead to biased results. It is hence of paramount importance to take this stylized fact into consideration when designing a fill probability model. The intuitive way of doing this is training the model on a data set of real limit orders. In practice, a trader could use her own trading history in order to take into consideration a cancellation tactics. Secondly, and most importantly, using real-life order flow allows one to build a model for aggressive limit orders, a task that is impossible to carry out when using a first passage time method. Last but not least, the raw size of the posted order becomes a key feature of the model as it obviously plays a significant role in a high-frequency setting, and the model can therefore learn the intricacies of its role over market's reaction.

\subsection{Model training and feature importance}

\subsubsection{Training pipeline}

We formulate the fill probability estimation problem as a binary classification problem with a suitable loss weighting procedure to account for censored data. For a data set with matrix representation $\mathbf{Z}:=(z_{ij})_{1\leq i \leq N, 1\leq j\leq d}$ of $N$ observations (rows) and $d$ features (columns), we denote by $\mathbf{y}:=(y_i)_{1\leq i\leq N}$ the vector of labels, where $y_i=0$ indicates that the $i$th order was not executed under time horizon $T$ and $y_i=1$ indicates this order was filled. Since the time horizon of interest is small, we discarded limit orders that were posted too far in the book in order to remove noisy observations. The market depth threshold was fixed at 20 basis points of the mid price for the digital asset data base and 5 price limits for the equity one.

In Section \ref{sec:non_parametric_analysis}, we have shown that the fill probability function presents smooth non-linear dependencies with respect to three new microstructure features, which makes the problem suitable for the training of neural networks. For this task, we use a feed-forward neural network with a sigmoid activation function for the output layer. We tested several architectures that all provided very similar results. For the results displayed in this work, we used 3 layers of 32 neurons with ReLU activation functions, and applied a 25\% dropout for each layer in order to improve generalization. We add other interpretable variables to the set of features introduced in Section \ref{sec:non_parametric_analysis}:

\begin{itemize}
    \item distance of the order to the best queue;
    \item best queue imbalance, defined as follows: if $q^b$ is the size of the best bid queue, and $q^a$ the size of the best ask queue, then the best bid-offer (BBO) imbalance is $\frac{q^b-q^a}{q^b+q^a}$;
    \item size of the order;
    \item bid-ask spread;
    \item signed limit order flow, which is defined as $\Delta_mQ^b-\Delta_m Q^a$, using the notations of Section \ref{sec:non_parametric_analysis};
    \item signed order flow and order flow imbalance using both addition of liquidity (limit order insertion) and removal of liquidity (cancellation of a pending order or transaction), computed over the last 50 events;
    \item signed traded volume and imbalance of traded volume, computed over the last 50 transactions. We adopt the liquidity taker's viewpoint, \textit{i.e.} if $V_{\text{bid}}$ and $V_{\text{ask}}$ are the traded volumes on the bid side and on the ask side over the last 50 transactions, then the signed traded volume is defined as $V_{\text{ask}} - V_{\text{bid}}$ and the traded volume imbalance as $\frac{V_{\text{ask}} - V_{\text{bid}}}{V_{\text{ask}} + V_{\text{bid}}}$;
    \item time elapsed since the last trade occurrence and the median duration of the last 50 trades, which could be characterized by the intensity of some self-exciting point process;
    \item  volatility defined from a high-frequency estimator based on the uncertainty zone model of \cite{robert2011new}, and computed on traded prices of a moving window of 50 trades.
\end{itemize}

We rescale the variables using the classical Box-Cox transformation followed by a $z$-score. The hyperparameter of the Box-Cox transformation is chosen such that the statistic of the Kolmogorov-Smirnov test versus the standard Gaussian is minimized.

The significant presence of right censoring in the data set leads us to consider a loss weighting methodology. Simply discarding orders that are cancelled before the time horizon $T$ from the data set would lead to a significant overestimation of the fill probability function. We use the “inverse-probability-of-censoring weighting” (IPCW) method to take care of this issue. This method is a well-known technique \cite{mark1993method} and was successfully used in many real-world problems, see \cite{vock2016adapting} and \cite{gonzalez2021stacked} for the details of the method and a comparative analysis. It was shown in \cite{satten2001kaplan} that an IPC-weighted version of the estimator of the survival function without censoring is equivalent to the Kaplan-Meier estimator, hence justifying the construction of this methodology. To our knowledge, our work is the first to apply IPCW to the training of a fill probability model.

Denote by $w_i$ the weight associated to observation $y_i$, $C_i$ the time of censoring which can be either a cancellation or other causes of censoring and $E_i$ the time of execution. The IPC weights are defined as follows

\begin{equation}
    w_i:=\frac{\mathds{1}_{\{\text{min}(E_i, T)<C_i\}}}{\mathbb{P}\big(C_i>\text{min}(E_i, T)\big)}.
\end{equation}

The high-frequency activity occuring mainly near the mid price, we introduce a dependence of the weights to the distance of placement of limit orders. Such a modification will increase the weight applied to orders that were posted near best prices and even more for orders which manage to stay in the book until the time horizon $T$. Concerning orders that are posted inside the bid-ask spread, we propose to condition on the aggressiveness index as defined in Equation \eqref{eq:aggressiveness_index} to take into account the high cancellation rate of less aggressive orders. The censoring survival function is computed using the Kaplan-Meier estimator of Equation \eqref{eq:kaplan_meier_estimator} but in this specific case, considering cancellation and right-censoring as death and execution as right-censoring.

The IPCW procedure deforms the fill probability in a similar way that the Kaplan-Meier function does by giving extra weights to orders that were not executed under the horizon or posted in highly censored configurations.

We focus on the bid side of BTC-USD and BNPP. We train the crypto model on 5 days --- from 2022-11-11 to 2022-11-15 --- and validate it on 2022-11-16. Note that we conducted the same test for other pairs and they all yielded similar results, which emphasizes the universality of the predictive power of the features we propose. For the equities, we train the model on 8 months ---from January $1^{\text{st}}$, 2017 to August $31^{\text{st}}$, 2017--- and validate it on 1.5 months ---from September $1^{\text{st}}$, 2017 to October $15^{\text{th}}$, 2017. The much longer calendar duration in that case compensates for the much lower trading activity of equities.

\subsubsection{Feature importance}

\begin{figure}[!htbp]
    \centering
    \subfloat[Contributions for passive orders $\delta > 0$, BTC-USD]{%
        \includegraphics[width=0.4\linewidth]{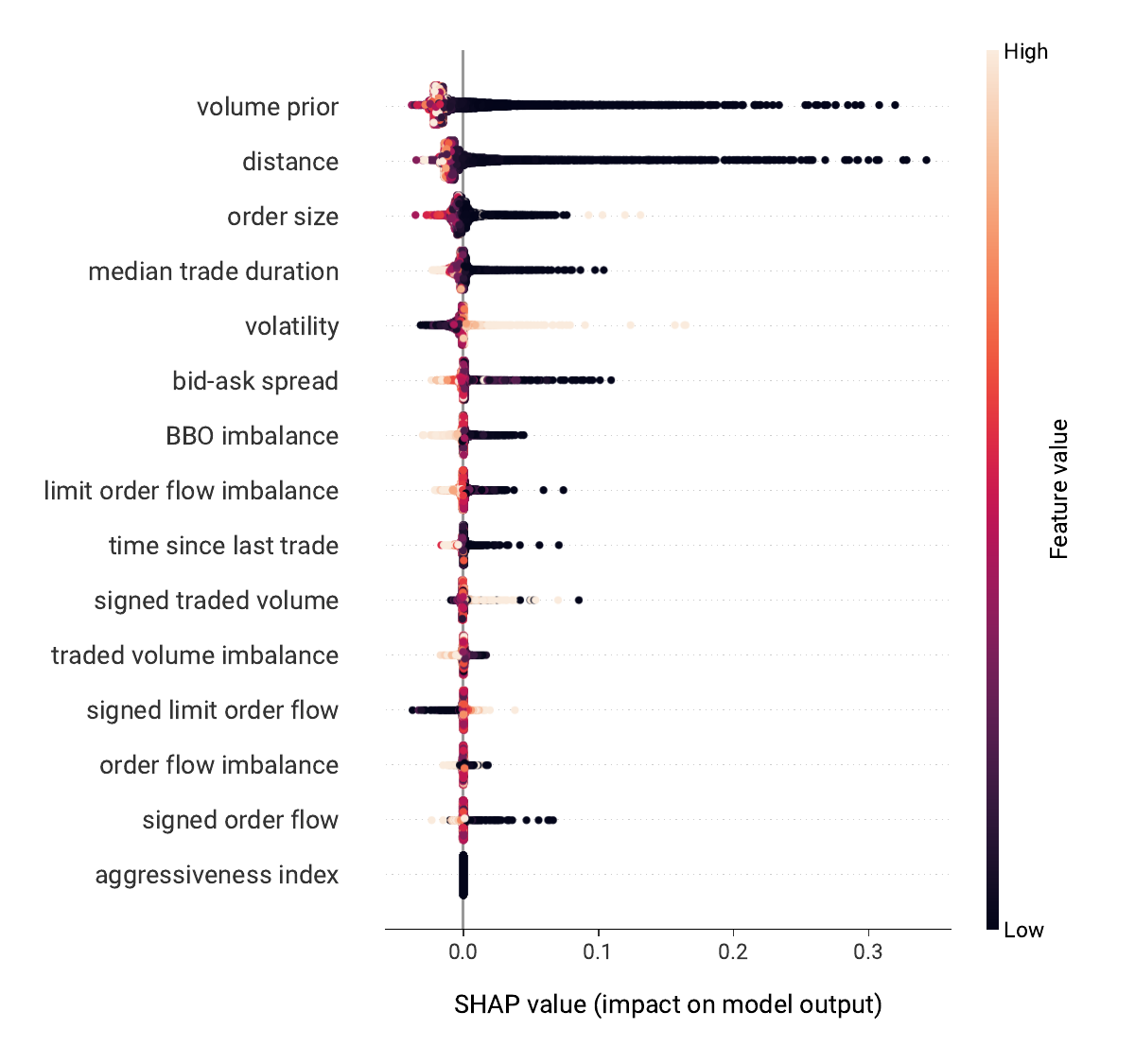}%
    }
    \subfloat[Contributions for passive orders $\delta > 0$, BNPP]{%
        \includegraphics[width=0.4\linewidth]{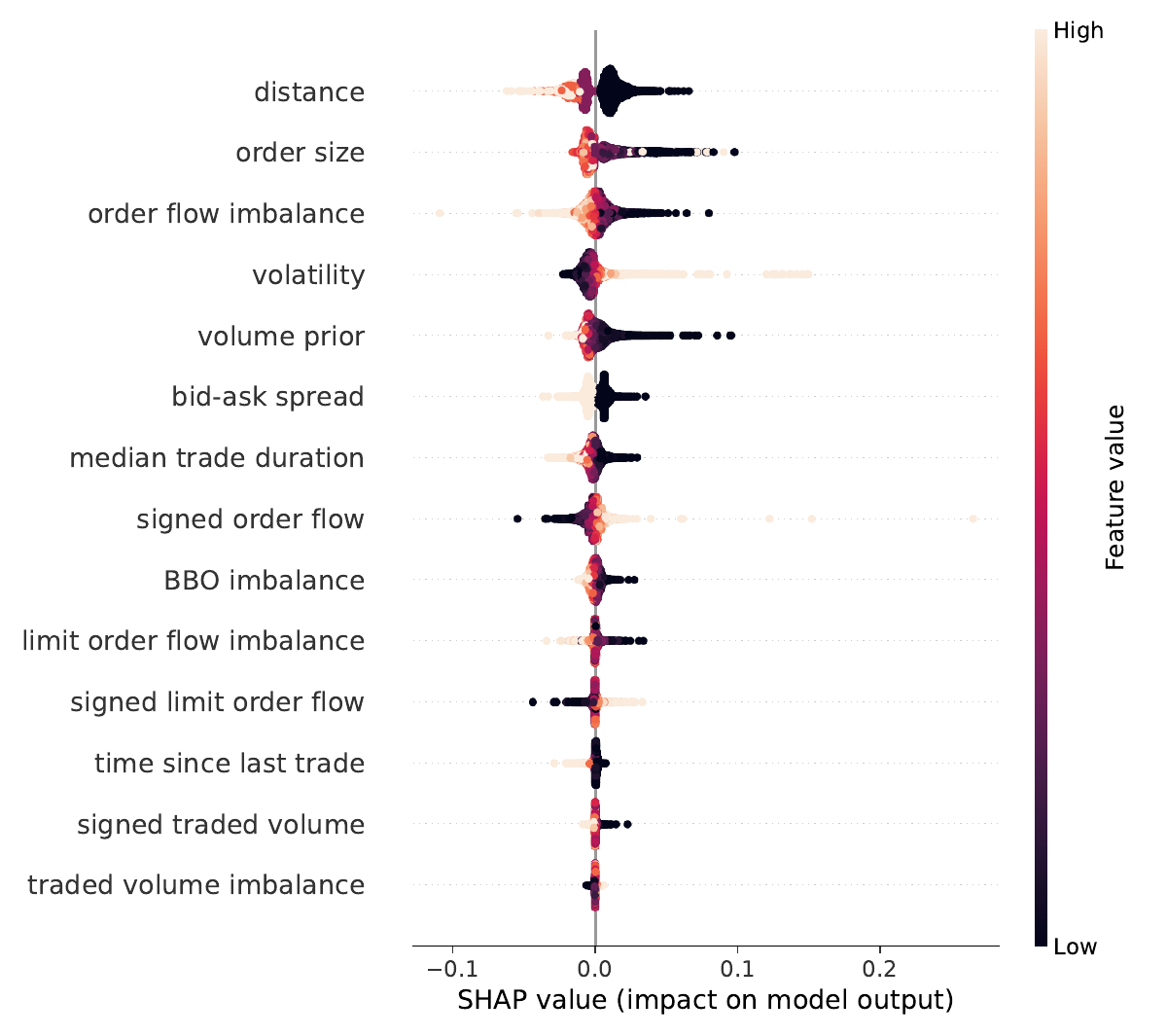}%
    }\\
    \subfloat[Contribution at best quote $\delta = 0$, BTC-USD]{%
        \includegraphics[width=0.4\linewidth]{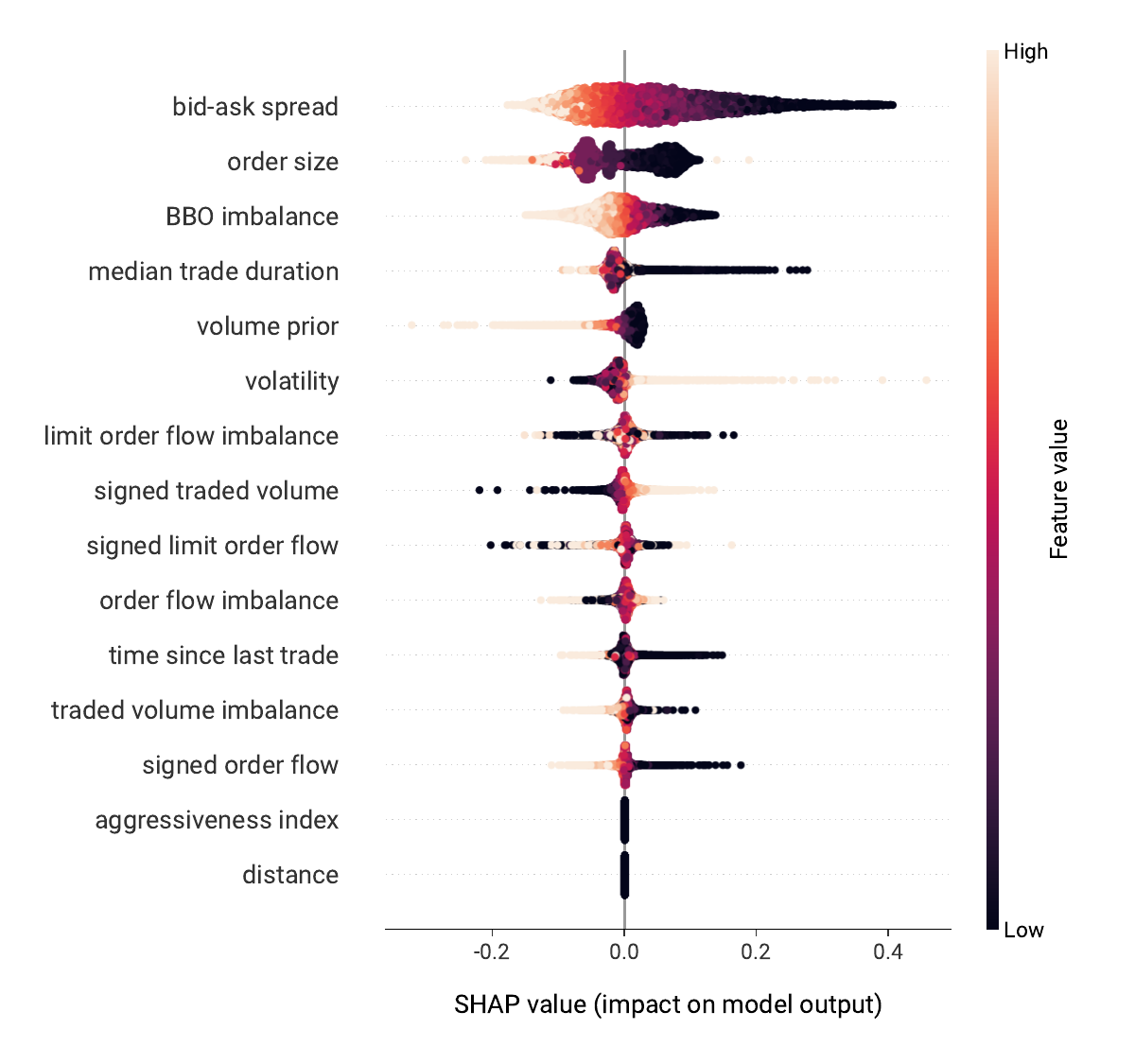}%
    }
    \subfloat[Contributions at best quote $\delta = 0$, BNPP]{%
        \includegraphics[width=0.4\linewidth]{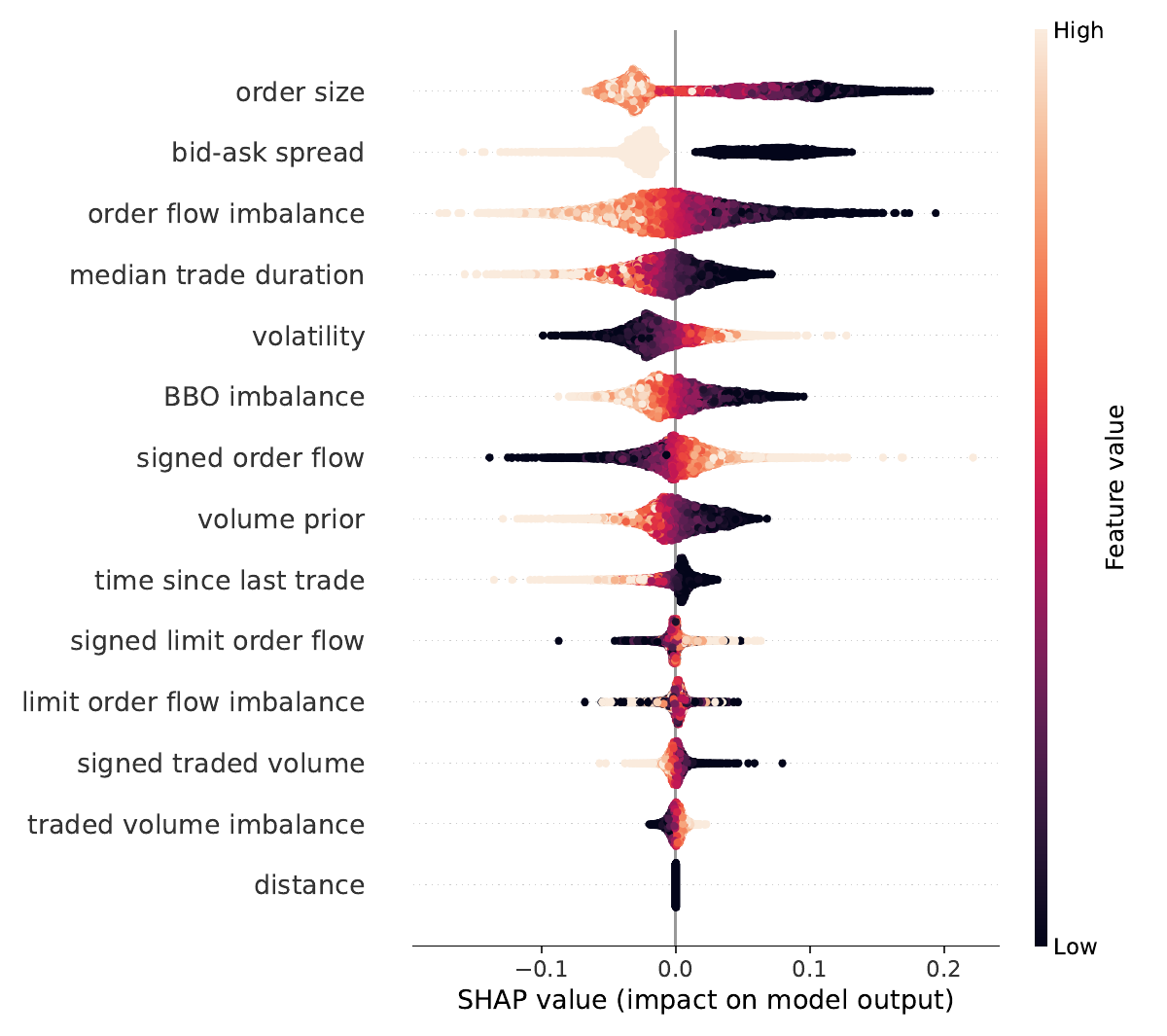}%
    }\\
    \subfloat[Contributions for aggressive orders $\delta < 0$, BTC-USD]{%
        \includegraphics[width=0.4\linewidth]{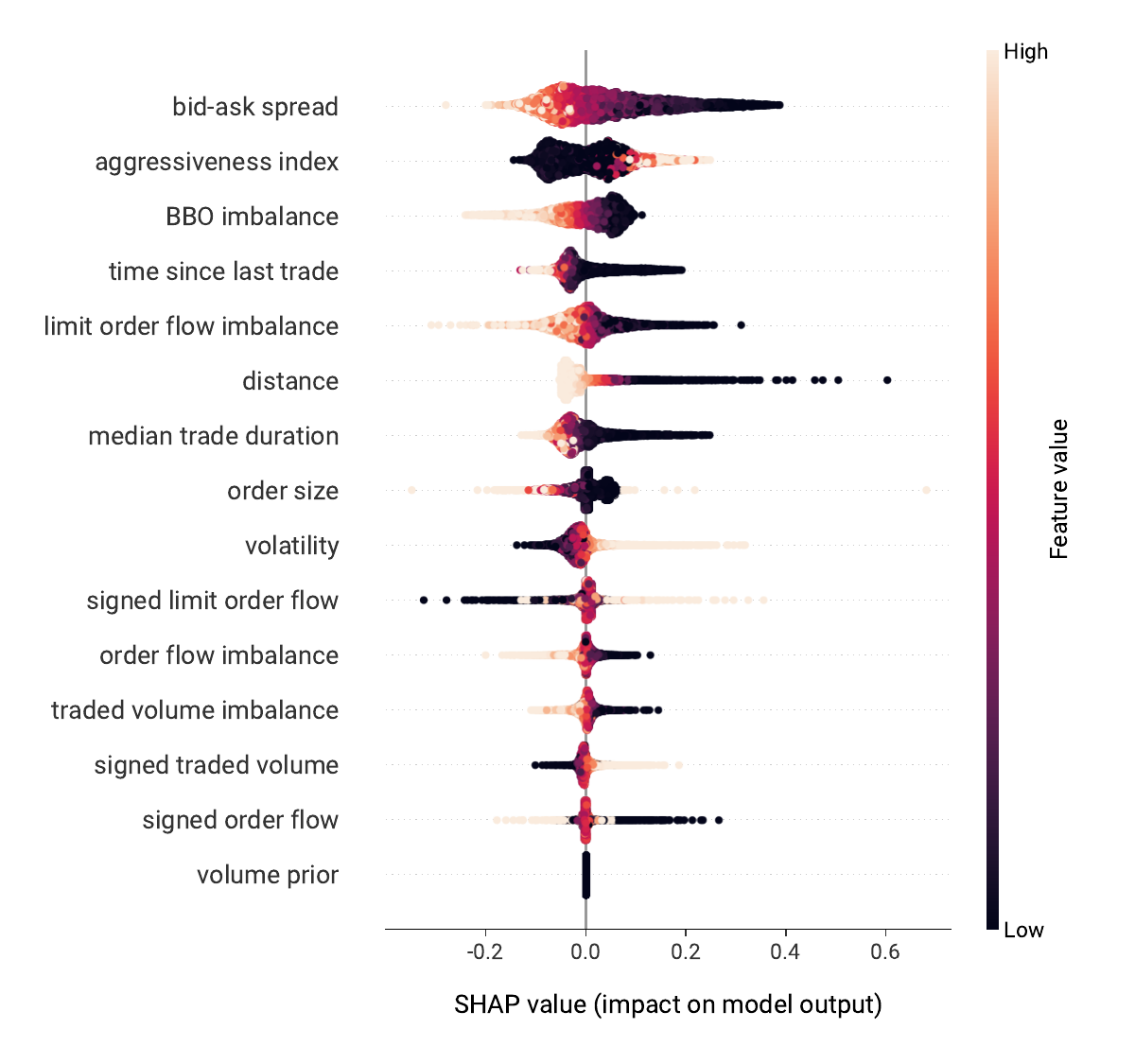}%
    }
    \subfloat[Contributions for aggressive orders $\delta < 0$, BNPP]{%
        \includegraphics[width=0.4\linewidth]{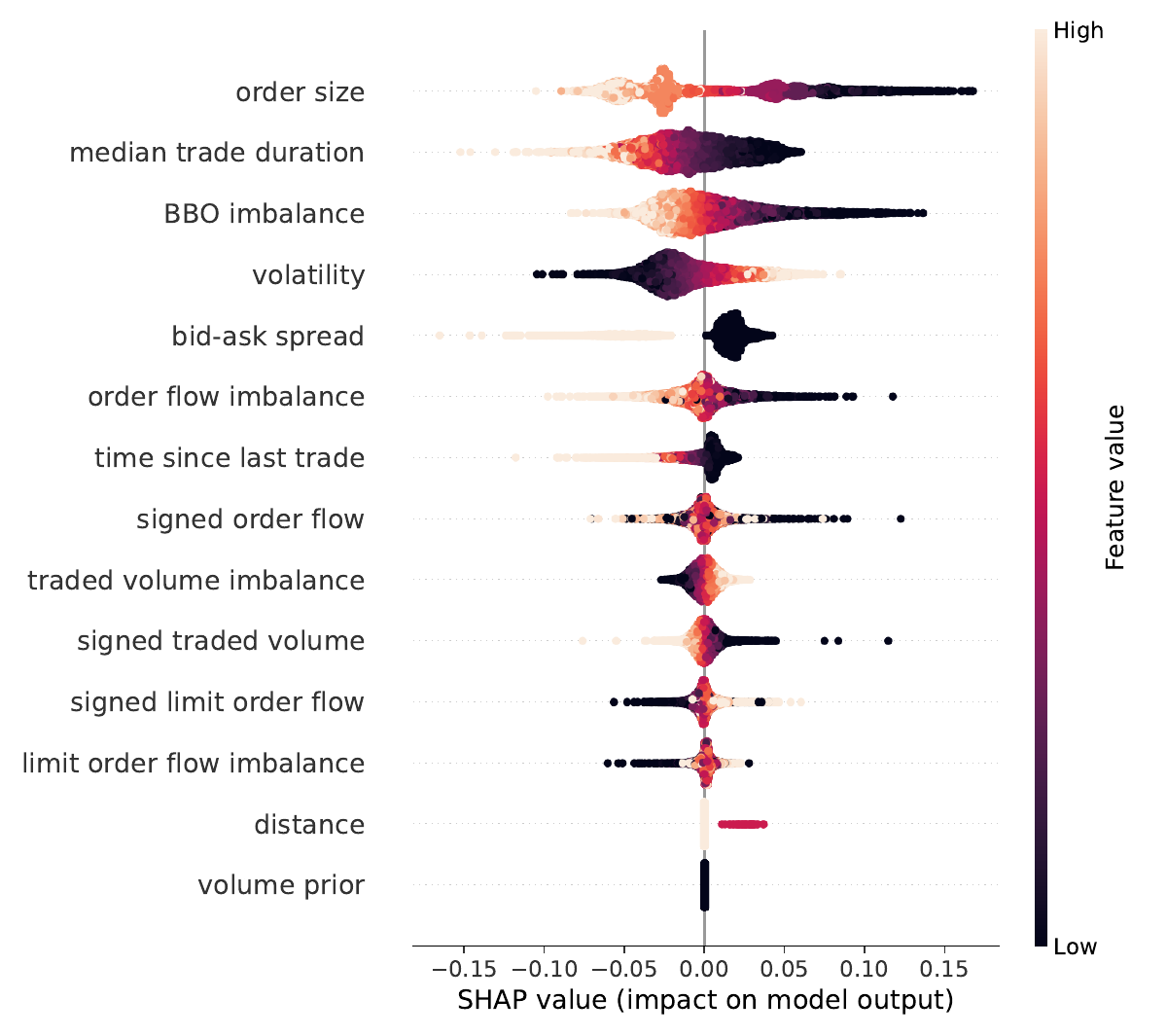}%
    }
    \caption{\textit{Feature importance} --- Market features contributions to the fill probability magnitude using Shapley values of 10,000 predictions, BTC-USD and BNPP, bid side.}
    \label{fig:shapley_fill_probability}
\end{figure}

We analyze the importance of features with Shapley values computed over out-of-sample observations using the SHAP library \citep{NIPS2017_7062}. The analysis is separated into three parts and brings insights about how the predictive power of features changes from order to order type, \textit{i.e.} for aggressive, at-touch, and passive orders. Shapley diagrams are displayed in Figure \ref{fig:shapley_fill_probability}.

\begin{enumerate}
    \item \textbf{Passively posting ($\delta>0$):} The distance and the order size are amongst the three most important features for both BTC-USD and BNPP. Interestingly the most important feature for the small tick asset is $V_{\text{prior}}$. Our understanding is that the distance alone is not sufficient to correctly characterize the priority in small tick order books because of their sparsity. For equal values of prior volume, different distances mean a different level of sparsity of the order book. If a small volume is quoted under a high distance $\delta$, other market participants are likely to quote new prices at smaller distances than $\delta$ under the time horizon. Based on this thought, we believe both variables are inseparable when it comes to fill probability computation for small tick assets. The prior volume is also important for the equity but it comes after the order flow imbalance and the volatility. 
    
    \item \textbf{Posting at the current best ($\delta=0$):} When posting at the current best queue, the bid-ask spread and the order size appear to be the most important features for both BTC-USD and BNPP. Interestingly, the importance of the BBO imbalance is smaller than that of the order flow imbalance for BNPP, which shows the importance of the dynamic features allowing the model to capture changes in the order flow.
    
    \item \textbf{Aggressively posting ($-\psi<\delta<0$):} When a new best queue is created, the bid-ask spread and the aggressiveness index are the most important features for the BTC-USD pair whereas for BNPP, the order size and the median trade duration are the most predictive ones. Since the spread of small tick assets is generally larger than one tick, the fill probability of an aggressive order is conditioned on the strength of the spread tightening it induces. Moreover, the limit order flow imbalance brings also an important contribution for the cryptocurrency pair, emphasizing the predictive power of this imbalance measure.
\end{enumerate}

\section{Application: optimal order placement}

\subsection{The cost function approach}

We concentrate on the tactical aspect of optimal trading. We consider an agent who aims at buying a small quantity $q$ of an asset under a fixed horizon $T$ that ranges from milliseconds to seconds. Here, a small quantity implies that the transaction volume will be managed by a single order, which can be either a market order or a limit order. We show that incorporating deep knowledge about the execution probability can enhance the decision-making process, and thereby improve performance.

\subsubsection{Framework and notations}

We place ourselves in the Implementation Shortfall optimization framework: the reference price of the execution algorithm is the initial mid price $p_0$ and the execution schedule must minimize the expected difference between the execution price (including fees) and this reference price. Before we detail the decision process, we need to introduce useful notations: denote by $\big(p_t^b\big)_t$, $\big(p_t^a\big)_t$, $(p_t)_t$,  $\left(\psi_t\right)_t$ respectively the best bid price, the best ask price, the mid price, and the bid-ask spread processes of the asset. The tick size, expressed in quote units, is denoted by $\alpha$, and we denote for $0\leq t\leq T$ the variation of the best price  $\bullet$ over $[0,t]$ by $\Delta p_t^\bullet:=p_t^\bullet-p_0^\bullet$, for $\bullet\in\{b, a\}$. The agent observes a market state vector $z\in\mathbb{R}^d$. 

We denote by $\varepsilon^-$ and $\varepsilon^+$ the taker and maker transaction fees respectively, such that $\varepsilon^->\varepsilon^+$, and define the fee factors $f^-:=1+\varepsilon^-$ and $f^+:=1+\varepsilon^+$. Note that in CEXs, the fixed transaction fees vary as a decreasing function of the traded volume which is often computed over a 30 days rolling window.

The agent has access to a limit order book and thus chooses between the two following tactics at the initial time 0.

\begin{itemize}
    \item \textbf{Immediate execution tactic:} The agent crosses the bid-ask spread by sending a marketable order to get immediate execution at price $p_0^a$. The cost of this tactic is deterministic and will be denoted by $\mathcal{M}$. We suppose that $q$ is sufficiently small such that the corresponding execution price is the best ask price $p_0^a$, \textit{i.e.} there is no immediate market impact.
    \item \textbf{Post and wait tactic (PW):} The agent posts a buy limit order at bid price $p_0^b-\alpha\,\delta$ and waits for its execution until the time horizon $T$. The parameter $\delta$ is the distance to the best bid price as a number of ticks. The lifetime of this order will be denoted by $L^{\delta,q}$ following the conventions introduced in Section \ref{sec:non_parametric_analysis}. Note that it is indexed by $\delta$ and $q$ since it is associated to an order of size $q$ placed at a distance $\delta$ to the best bid. We will denote by $F_T^{\delta,q,z}:=\mathbb{P}\left(L^{\delta,q}\leq T|Z=z\right)$ its fill probability within time horizon $T$ conditionally on a market state $z$. The cost function of the post and wait tactic is random and will be denoted by $\mathcal{W}(T,\delta,q)$.
\end{itemize}

As full execution is not guaranteed in the PW case, the agent will send a marketable order for immediate execution at the end of the period if the limit order is not filled by then, which will incur additional transaction costs in the case of adverse price moves. Both the fill probability and the clean-up cost increase with the volatility: the agent needs to find a trade-off between certainty of execution and management of the clean-up cost induced by market risk.

Henceforth, we remove the size factor $q$ from the cost functions for the sake of clarity; in addition,  $\mathbb{E}_z[.]$ corresponds to the conditional expectation $\mathbb{E}[.|z]$.

The expected execution costs of the two tactics are written as

\begin{equation}
    \mathbb{E}_z\left[\mathcal{M}\right]=\mathcal{M}=f^-p_0^a-p_0,
\end{equation}
and

\begin{align}
    \mathbb{E}_z\left[\mathcal{W}(T,\delta,q)\right]&=F_T^{\delta,q,z}\big(f^+(p_0^b-\alpha\,\delta)-p_0\big)+(1-F_T^{\delta, q, z})\,\mathbb{E}_z\left[\big(f^-p_T^a-p_0\big)|L^{\delta,q}>T\right].
\end{align}

Note that the case of a partial execution is not taken into account since it is an unlikely occurence given that we consider small orders. Indeed, we checked empirically that a negligible proportion of orders were only partially executed under the time horizon $T$ that we have chosen for the experiment.

The expected cost reduction if the agent chooses the post and wait tactics over immediate execution, denoted by $\mathcal{S}$, is defined as

\begin{equation}
    \mathcal{S}(T,\delta,q,z)=\mathbb{E}_z\left[\mathcal{M}-\mathcal{W}(T,\delta,q)\right].
\end{equation}

Using elementary calculus, we find

\begin{align} \label{eq:saved_cost_function}
    \mathcal{S}(T,\delta,q,z)&=\underbrace{F_T^{\delta,q,z}\big(f^-p_0^a-f^+(p_0^b-\alpha\delta)\big)}_{\textbf{(a)}}-\underbrace{(1-F_T^{\delta,q,z})\,f^-\mathcal{V}_T^{\delta,q,z}}_{\textbf{(b)}},
\end{align}
where

\begin{equation}
    \mathcal{V}_T^{\delta,q,z}:=\mathbb{E}_z\left[\Delta p_T^a\Big|L^{\delta,q}>T\right]
\end{equation}
is the expected price variation of the best ask price over the period, conditionally on the non-execution of the pending order. To  understand the role played by all variables better, we decomposed Equation \eqref{eq:saved_cost_function} into two parts depending on the fate of the order:

\begin{itemize}
    \item \textbf{(a)} if the limit order is fully executed, the agent saves the spread between the initial net of fees best ask price at which an immediate liquidity taking would have occurred and the net of fees bid price of the filled order;
    \item \textbf{(b)} in the case of a non execution, the agent incurs a transaction cost that may be greater than if an immediate execution had been chosen at the beginning. This clean-up cost is unknown at the beginning of the period and is characterized by the function $\mathcal{V}$.
\end{itemize}

The function $\mathcal{V}$ exhibits significant sensitivities to many variables such as realized volatility. This is intuitive, considering that the cleanup cost inherently reflects measures of volatility. To give additional intuition about the behaviour of this function, note that we expect it to behave as a non-increasing function of the total volume pending at better prices than the price of the posted order. Indeed, if the order is posted at the current best queue and does not get filled under the time period, it indicates that the market may have moved in the other direction thus inducing additional costs. This sensitivity is even stronger with aggressiveness. If the order is posted far from the best price, a non-execution does not necessarily indicate an adverse move of the opposite best price. Given all this remark, it is apparent that treating the clean-up cost as a constant in the objective of minimization would be simplistic. 

Under suitable regularity conditions, the function $\mathcal{S}$ can be maximized over the set of admissible distances $\mathcal{A}_{\psi_0}:=\{\delta\in\mathbb{Z},\,\delta>-\psi_0\}$ to find the optimal order placement strategy at fixed $T$ and $q$, leading to the following optimization problem
\begin{equation}
    \delta^*\,=\,\underset{\delta\in\mathcal{A}_{\psi_0}}{\text{argmin}}\,-\mathcal{S}(T,\delta,q,z).
\end{equation}

The unicity of the maximum is a challenge itself since we are dealing with highly non-linear dependencies that are inferred with a neural network. We therefore put that question aside for future investigation.

\subsubsection{A toy model} \label{subsubsec:toy_model}

Before diving into the estimation of $\mathcal{V}$, let us introduce a simplified version of the order placement model with an exponential fill probability function as specified in \cite{avellaneda2008high}, \cite{laruelle2013optimal}. For the sake of clarity, we set $f^-=f^+=1$ and we consider the modified distance $\delta^a:=\psi_0+\delta$ of the order the best ask price at time 0, expressed in ticks. We set $\mathcal{V}$ constant and for all $\delta^a\geq1$,
\begin{equation}
    F^{\delta}=Ae^{-k\delta^a}.
\end{equation}

The expected saved cost function writes:
\begin{equation}\label{eq:saved_cost_function_toy_model}
    \mathcal{S}(\delta)=Ae^{-k\delta^a}\delta^a-\left(1-Ae^{-k\delta^a}\right)\mathcal{V}.
\end{equation}

Setting the condition $k(1+\mathcal{V})\leq 1$ and differentiating with respect to $\delta^a$, we obtain the following optimal distance of placement:
\begin{equation} \label{eq:toy_model_optimal_distance}
    \delta^{a,*}:=\frac{1}{k}-\mathcal{V}
\end{equation}
and the associated maximum of the saved cost function:
\begin{equation}
    \mathcal{S}(\delta^{a,*})=\frac{A}{k}e^{k\mathcal{V}-1}-\mathcal{V}.
\end{equation}

Equation \ref{eq:toy_model_optimal_distance} indicates that in this simple framework, the optimal distance should scale linearly with respect to the expected adverse price move $\mathcal{V}$, and inversely with respect to the decay rate of the fill probability. This model will later be used as a benchmark for assessing the performance of the full execution algorithm.

\subsubsection{The effect of CEXs fee policy on the strategy}

When trading in CEXs, agents may face very different transaction fees depending on their monthly turnover. We build an example to illustrate the strong sensitivity of the strategy to the fee policy.

\begin{figure}[!htbp]
    \centering
    \includegraphics[width=0.5\linewidth]{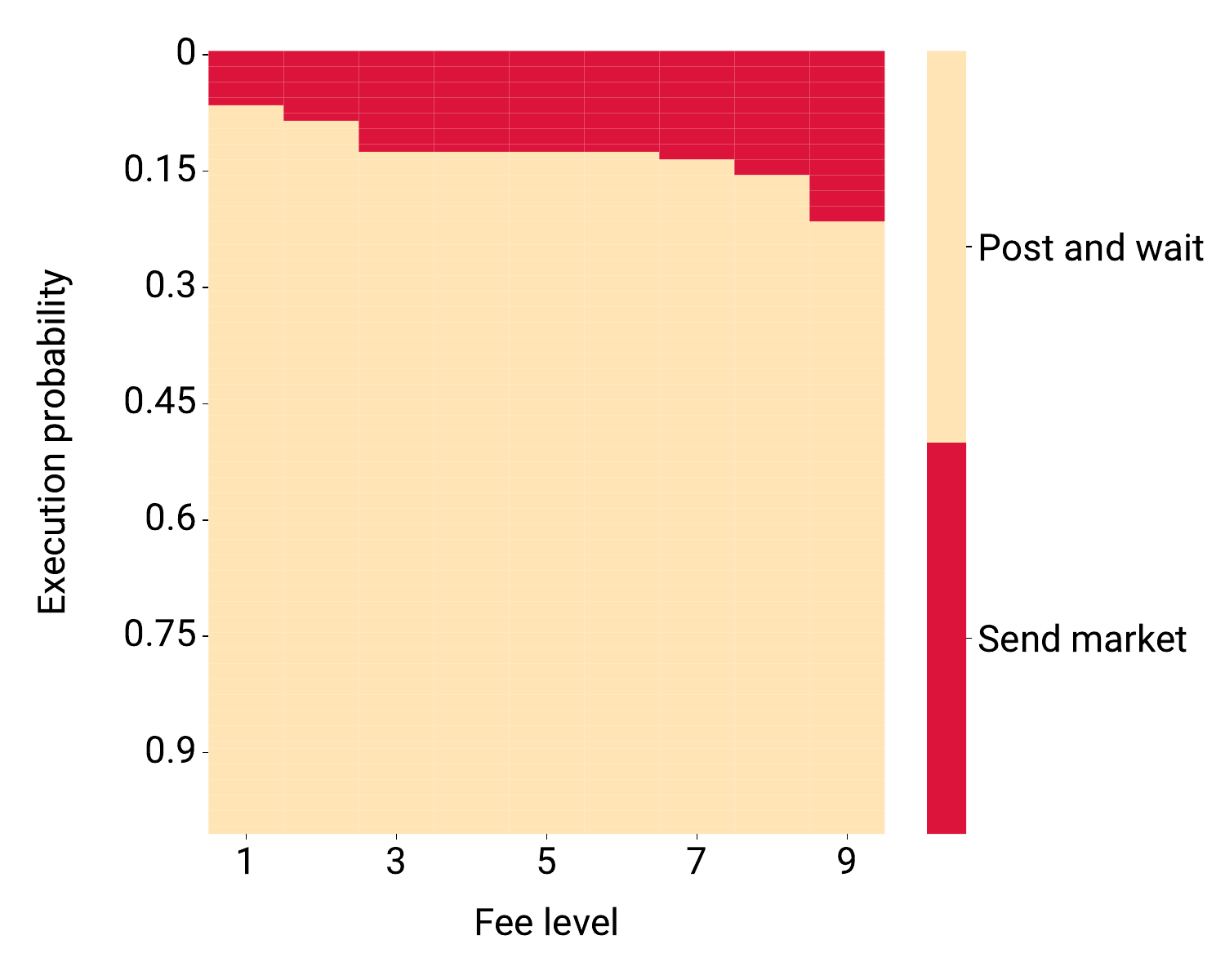}
    \caption{\textit{Decision boundary} --- Decision map in the practical example as a function of fee level and execution probability. Fee levels are displayed in Table \ref{tab:fee_policy}.}
    \label{fig:saving_cost_vs_fee_level}
\end{figure}

We consider an agent who posts a buy limit order in the book at the current best bid price $p_0^b=19,999.50$ USD and a bid ask spread of $1.00$ USD. Suppose that the model predicts a clean-up cost $\mathcal{V}=2.00$ USD (which would correspond to an annualized volatility of approximately 56\%). We present the trading fee policy of Coinbase in Table \ref{tab:fee_policy}. In Figure \ref{fig:saving_cost_vs_fee_level}, we display the decision map of the execution algorithm as a function of the fee level and the execution probability. We clearly observe that for a fixed execution probability, the fee level has a strong impact over the optimal decision.

\begin{table}[!ht]
    \caption{\textit{Trading fees} --- Coinbase fee policy of spot trading on April 2023.}
    \begin{center}
        \begin{tabular}{cccc}
            \toprule
            \toprule
            \textbf{Level} & \textbf{Monthly turnover} & $\varepsilon^-$ & $\varepsilon^+$\\
            \midrule
            1 & $[\$0, \$10\text{K})$ & 0.006 & 0.004 \\
            2 & $[\$10\text{K}, \$50\text{K})$ & 0.004 & 0.0025 \\
            3 & $[\$50\text{K}, \$100\text{K})$ & 0.0025 & 0.0015 \\
            4 & $[\$100\text{K}, \$1\text{M})$ & 0.002 & 0.001 \\
            5 & $[\$1\text{M}, \$15\text{M})$ & 0.0018 & 0.0008 \\
            6 & $[\$15\text{M}, \$75\text{M})$ & 0.0016 & 0.0006 \\
            7 & $[\$75\text{M}, \$250\text{M})$ & 0.0012 & 0.0003 \\
            8 & $[\$250\text{M}, \$400\text{M})$ & 0.0008 & 0 \\
            9 & $[\$400\text{M}, \infty)$ & 0.0005 & 0 \\
            \bottomrule
        \end{tabular}
    \end{center}
    \label{tab:fee_policy}
\end{table}

\subsubsection{Estimation of the expected adverse price move $\mathcal{V}$}

\begin{figure}[!htbp]
    \centering
    \includegraphics[width=0.5\linewidth]{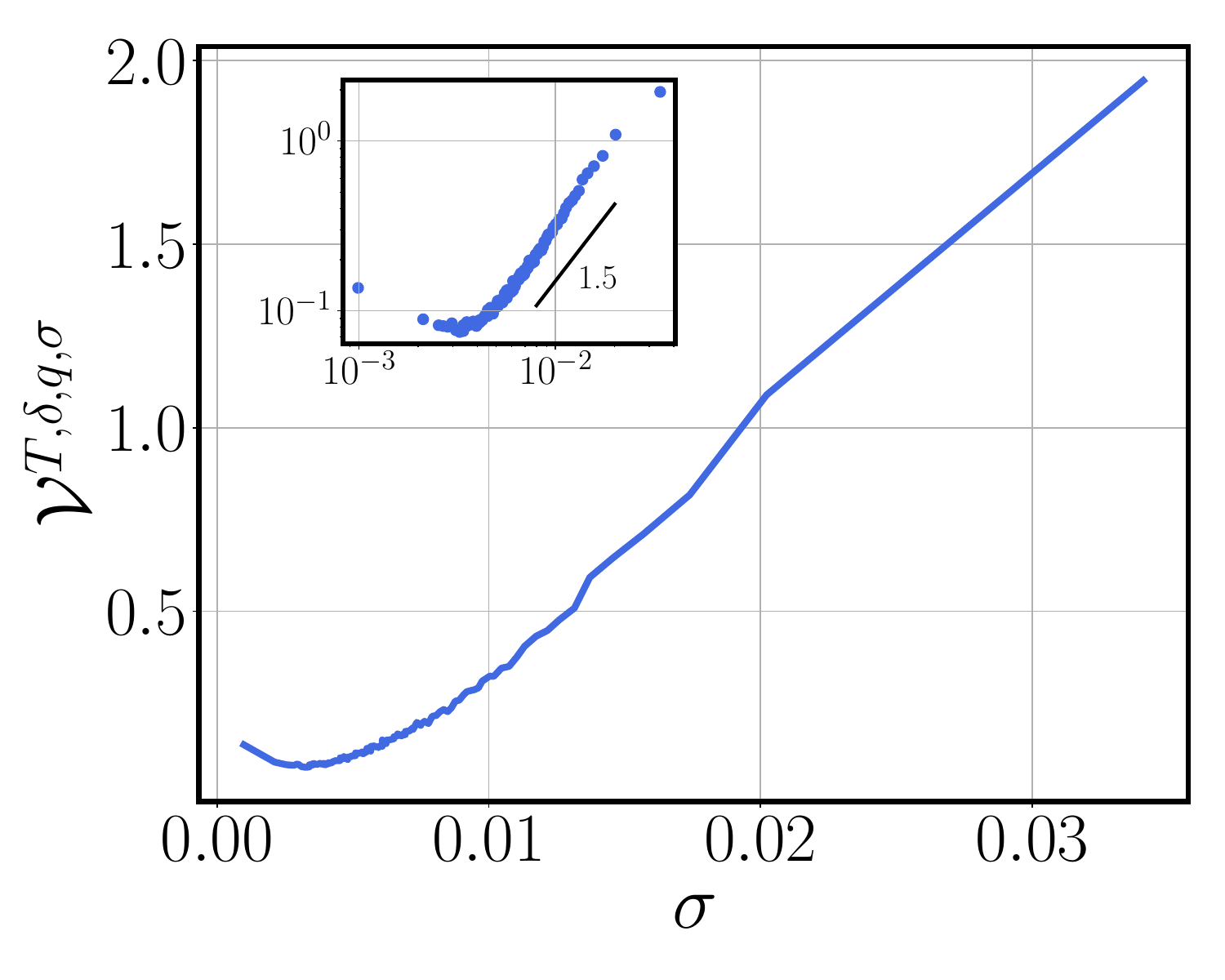}
    \caption{\textit{The clean-up cost as a measure of market risk} --- Empirical estimate of $\mathcal{V}$, in USD, as a function of realized high-frequency volatility $\sigma$, expressed in \%/trade, for BTC-USD.}
    \label{fig:clean_up_cost_volatility}
\end{figure}

We propose a simple methodology to estimate the clean-up cost function $\mathcal{V}$. Using the level 3 data we track the best ask price dynamics after each order's insertion and record its variation over the time window $[0, T]$ when the order is not executed. We remove the orders that were cancelled, executed or censored before $T$ since our only interest is in the events $\{L^{\delta,q}>T\}$. By doing so, we are able to build feature buckets and compute the average price move per bucket. An example of the shape for the estimator of $\mathcal{V}$ with respect to the realized volatility $\sigma$ is presented in Figure \ref{fig:clean_up_cost_volatility}. We observe a smooth dependence, indicating that the function $\mathcal{V}$ shares analogous properties with the fill probability function and can be estimated using a neural network model, and, once again, with a simple architecture. This measure is closely related to the volatility, which makes it much more predictable than the raw price moves.

We train a neural network with the same architecture as the one of the fill probability model, but with a linear activation function in the output layer. Using both trained NNs, the saved cost function $\mathcal{S}$ is finally estimated using Equation \eqref{eq:saved_cost_function}. Further analysis of the importance of each variable in the decision making process, can be found in Appendix \ref{sec:appendix}, Figure \ref{fig:shapley_saved_cost}. In addition to the key features that we described for the fill probability function, the limit order flow imbalance turns out to play a major role for cryptos in the estimation of the saved cost function in the three considered configurations: passively posting, posting at the current best queue and aggressively posting.

\subsection{Backtest of the order placement router}

Historical backtests are often misleading for many practical reasons such as the absence of market impact that would be undoubtedly caused by the algorithm on a real market. For example, the insertion of a new limit order in the first queue would adversely modify the best queues imbalance and negatively impact the strategy execution outcome. Considering infinitesimal sizes for orders does not help that much considering it will not be the case once the algorithm is sent to production. This is even more true for aggressive orders since they cause an immediate spread narrowing, thus inducing much more reaction from the market. The absence of market impact inflates the true out of sample performance, and another solution needs to be proposed for high frequency execution algorithms.

\subsubsection{Towards an impact-adjusted backtest}

We select real limit orders that were posted in the book and compute their expected saved costs $\mathcal{S}$. Each order should either be filled under the horizon $T$ or left in the book at least for a time $T$. The characteristics of the selected orders should reflect those of the orders that are posted by the strategy. For example, if our execution tactic posts orders with sizes ranging from 100 USD to 1,000 USD, and with distances smaller than 5 basis points, then the data set should be composed of orders with respect to these constraints. For each limit order in this test set, the sign of $\mathcal{S}$ will indicate whether the execution algorithm would have effectively posted the order, \textit{i.e.} $\mathcal{S}>0$, or opted for immediate execution instead, \textit{i.e.} , \textit{i.e.} $\mathcal{S}<0$. This step leads to an hypothetical decision $d=\mathds{1}_{\mathcal{S}>0}$. Then, one of the following three outcomes is observed at the time horizon $T$.

\begin{enumerate}
    \item The limit order was executed. The true optimal decision is $\hat{d}=1$.
    \item The limit order was not filled under $T$, but the best ask price variation is negative, hence leading to an improved execution price. The true optimal decision is $\hat{d}=1$.
    \item The limit order was not filled under $T$, and the best ask price variation is positive, hence leading to extra transaction costs. The true optimal decision is $\hat{d}=0$.
\end{enumerate}

This labelling procedure enables us to deduce binary classification metrics to assess the performance of the model in making the right decision, by comparing the predicted $d$ with the observed $\hat{d}$. It is important to note that by proceeding so, we are able to test the decision-making algorithm, but not the effectiveness of the optimal distance of insertion.

\begin{figure}[!htbp]
    \centering
    \includegraphics[width=0.8\linewidth]{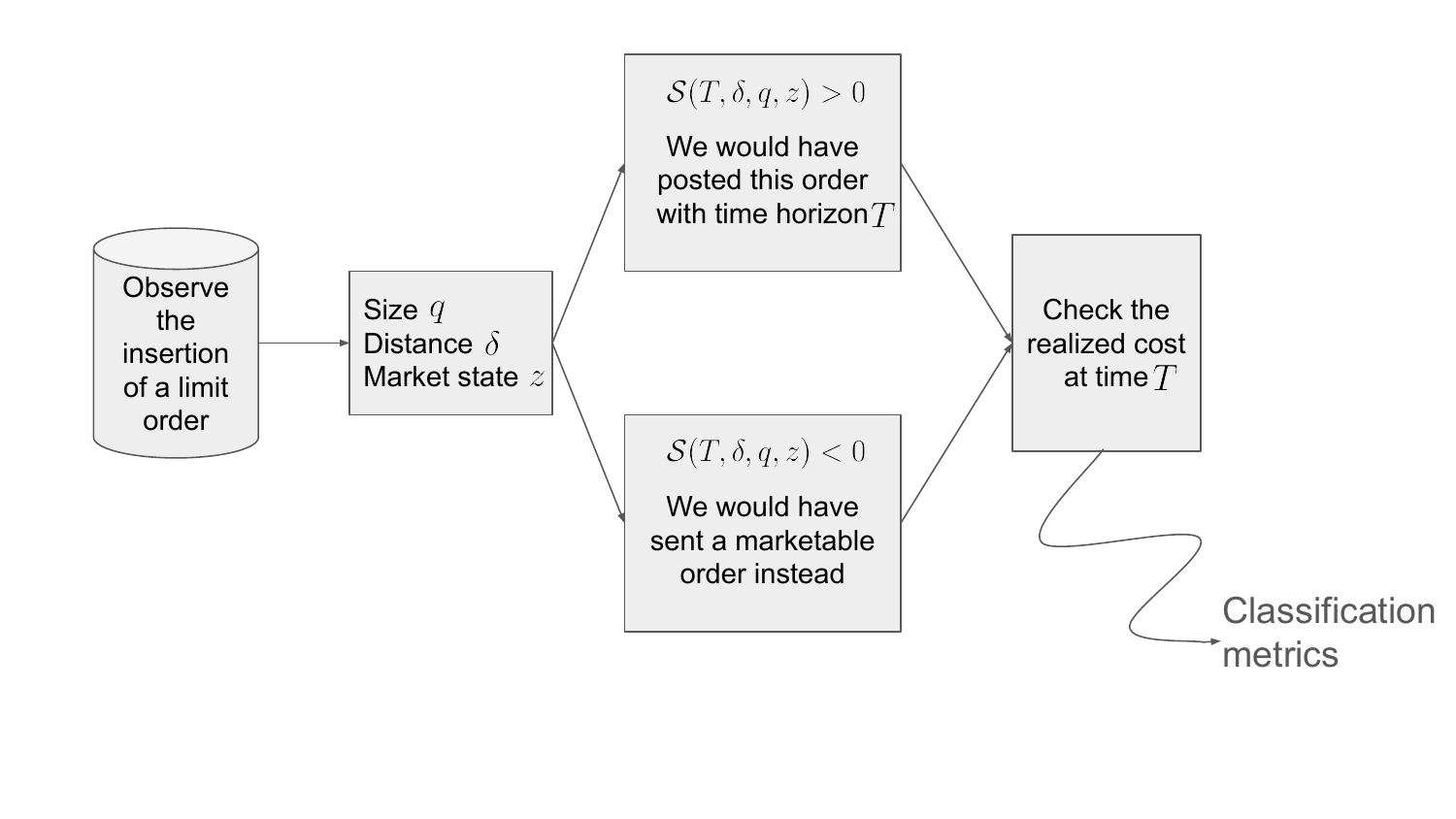}
    \caption{\textit{Impact-adjusted backtest procedure} --- Diagram of the methodology for performance evaluation of a tactical execution algorithm.}
    \label{fig:backtest_diagram}
\end{figure}

The backtest procedure is illustrated in the diagram of Figure \ref{fig:backtest_diagram}.

\subsubsection{Experiment setting}

For the CEXs data base, both fill probability and clean-up cost models are trained on 5 days, validated on two days, and tested on the following week, representing 3 test periods over the month. The equity model is trained on 8 months, validated on 1.5 months, and tested on 2.5 months.

We now describe the three models that will be tested and compared below. The chosen fee policy for the crypto trading algorithm is the level 9, \textit{i.e.} $\varepsilon^-=5$ basis points, $\varepsilon^+=0$.

\begin{itemize}
    \item \textbf{Model I:} The toy model introduced in \ref{subsubsec:toy_model} provides a closed form expression of the saved cost for any distance of placement $\delta$, see Equation \eqref{eq:saved_cost_function_toy_model}. The fill probability parameters $A$ and $k$ are estimated by fitting an exponential form on the Kaplan-Meier function considering cancellation as right-censoring and $\mathcal{V}$ is computed as the average of best ask price moves observed at horizon for every limit orders in the training set that are not executed. A more sophisticated estimation procedure could be adapted, for example by taking into account the intraday seasonality of market activity.
    \item \textbf{Model II:} The fill probability function used for this benchmark is the neural network model described in Section \ref{sec:fill_probability_model}, and $\mathcal{V}$ is a constant computed as in the exponential toy model benchmark. Thus, the only difference with the full model resides in the estimation of $\mathcal{V}$.
    \item \textbf{Model III:} Combination of the two neural network models for the fill probability and the expected best price move at horizon.
\end{itemize}

The results are displayed in Table \ref{tab:model_performance_decision_making}, which reports the respective performance of the three models for BTC-USD and BNPP. We observe a clear improvement of the decision-making process with the use of handcrafted features and non-linear models. It demonstrates that a simple and interpretable neural network architecture fed with well-chosen features may be sufficient to design decent tactical execution algorithms. Furthermore, we observe that using a state-dependent market risk $\mathcal{V}$ instead of a constant one seems to be crucial, as the F-score for both asset classes drastically improves. Interestingly, the algorithm performs much better on BTC-USD than on the BNPP stock. One possible interpretation is that BTC-USD is more predictable than BNPP.

\begin{table}[!ht]
    \caption{\textit{Impact-adjusted backtest} --- Performance metrics of the backtest of three order placement trading engines}
    \begin{center}
        \begin{tabular}{clccc}
            \toprule
            \toprule
            \textbf{Asset}         & \textbf{}           & \textbf{Precision}            & \textbf{Recall}    & \textbf{F-score}\\
            \midrule
                                   & I        & 0.21               & 0.99       & 0.34  \\
            BTC-USD                  & II       & 0.27               & 0.66       & 0.38    \\
                                   & III      & 0.37               & 0.81       & \textbf{0.51}  \\
            \midrule
                                   & I            & 0.11           & 0.05          & 0.07        \\
            BNPP                  & II           & 0.10           & 0.21          & 0.14        \\
                                   & III          & 0.20           & 0.54          & \textbf{0.30}        \\
            \bottomrule
        \end{tabular}
    \end{center}
    \label{tab:model_performance_decision_making}
\end{table}

\subsubsection{Analysis of the optimal distance for small tick assets}

\begin{figure}[!htbp]
    \centering
    \includegraphics[width=0.4\linewidth]{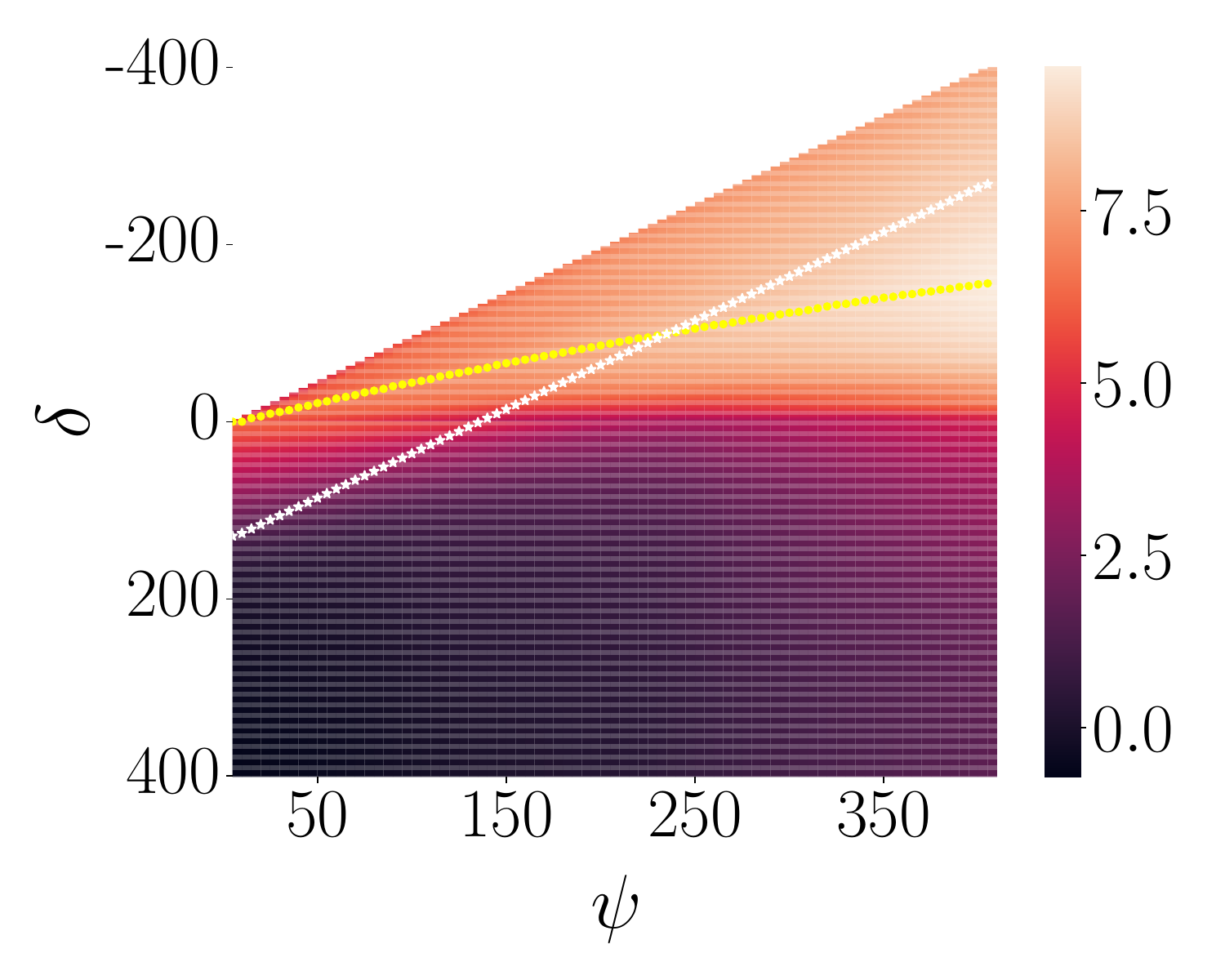}
    \caption{\textit{Optimal distance of placement} --- Example of the optimal distance policy $\delta^*$, as a function of the bid-ask spread $\psi$, BTC-USD setting fees to the level 9 of Table \ref{tab:fee_policy}. Both the distance $\delta$ and the bid-ask spread $\psi$ are expressed in number of ticks. The color bar represents the values taken by the expected saved cost $\mathcal{S}$. The optimal distance $\delta^*$ of model I given by Equation \eqref{eq:toy_model_optimal_distance} is represented by a white $\star$ and the optimal distance from model III is represented by a yellow $\circ$.}
    \label{fig:optimal_distance_2d_with_fees}
\end{figure}

\begin{figure}[!htbp]
    \centering
    \includegraphics[width=0.4\linewidth]{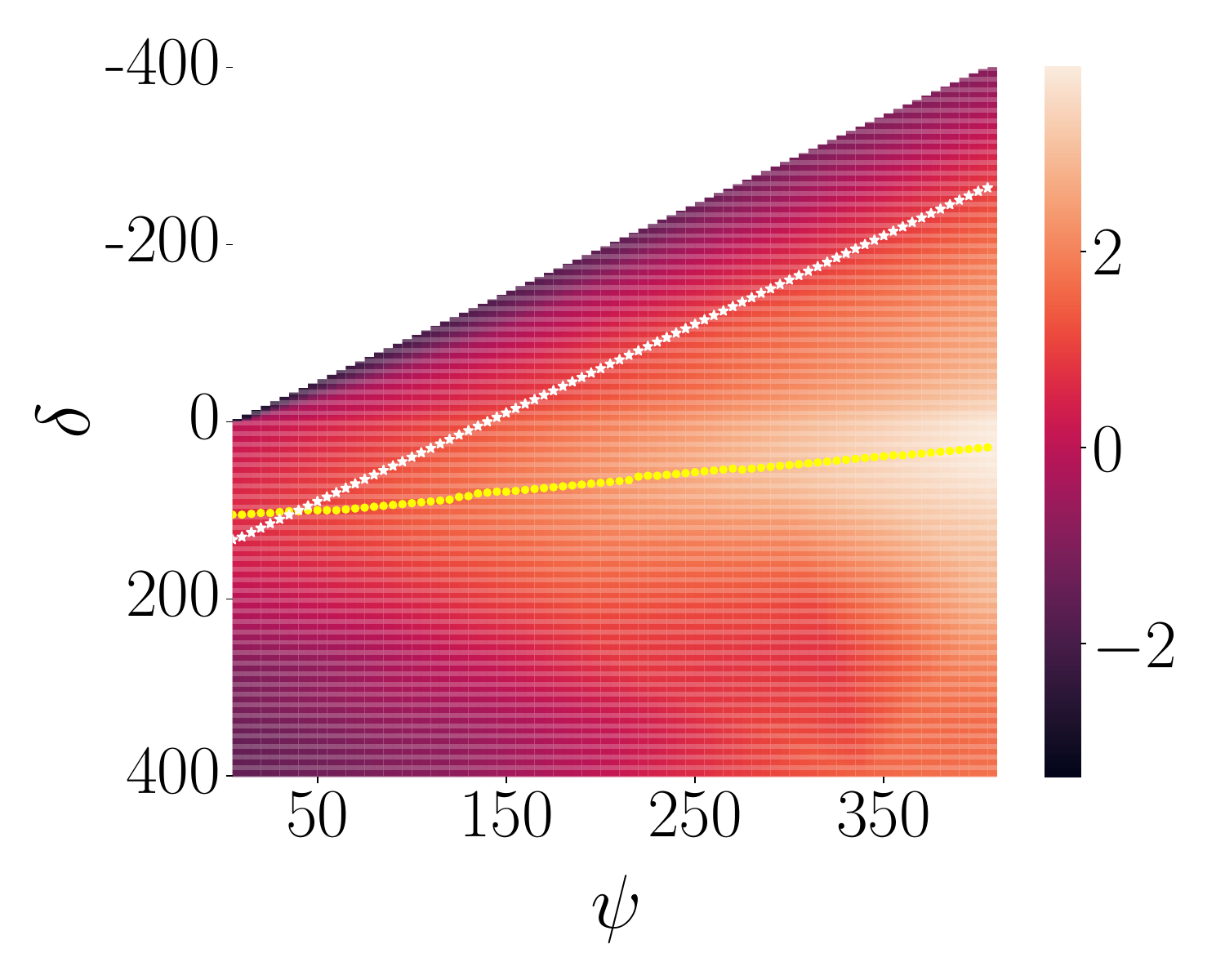}
    \caption{\textit{Optimal distance of placement} --- Example of the optimal distance policy $\delta^*$ as a function of the bid-ask spread $\psi$, BTC-USD, setting the fees to zero. Both the distance $\delta$ and the bid-ask spread $\psi$ are expressed in number of ticks. The heatmap represents the values taken by the expected saved cost $\mathcal{S}$. The optimal distance $\delta^*$ of model I given by Equation \eqref{eq:toy_model_optimal_distance} is represented by a white $\star$ and the optimal distance from model III is represented by a yellow $\circ$.}
    \label{fig:optimal_distance_2d_without_fees}
\end{figure}

Using model III, we compute the optimal distance of placement for BTC-USD and illustrate the behaviour of the algorithm with two cases. The first one is displayed in Figure \ref{fig:optimal_distance_2d_with_fees} and represents a heatmap of the expected saved cost $\mathcal{S}$ as a function of the bid-ask spread and the distance of placement, with the Level 9 fee policy corresponding to 5 bps of taker fees and no maker fees. The heatmap is computed using features that are observed at a random point in time. We observe that the algorithm tends to be aggressive and to quote inside the spread. It indicates that it exclusively focuses on maximizing the fill probability and saving the taker fees, even though the best way to do it is to narrow the spread and reduce the price discount of the limit order. The second illustration is displayed in Figure \ref{fig:optimal_distance_2d_without_fees} and is also a heatmap of the expected saved cost $\mathcal{S}$ computed at the same point in time as the previous one, but assuming there are no fees. We observe that the algorithm quotes deeper in the book than in the previous case. To wrap this up, the algorithm tends to post aggressive orders when the taker - maker fee gap $\varepsilon^--\varepsilon^+$ is large, pushing the optimal distance towards $-\psi$ as this gap increases. This result is particularly interesting as it demonstrates that CEXs fee policies tend to push agents to be more aggressive when it comes to executing fast. We believe this brings new elements of understanding of the impact of the fee policy on trading activity and agents behaviours in CEXs. Finally, we see that the optimal distance of placement given by the full model seems to be linear or at least sub-linear in the bid-ask spread, which is consistent with the prediction of the toy model we introduced.

\subsubsection{Penalizing aggressiveness with latency risk: a practitioner viewpoint}

In the case of small tick assets, we observed the resulting optimal order placement policy can place limit orders inside the spread. While posting a limit order in the spread guarantees a higher execution probability, this action brings a risk to the table: the instantaneous volatility of the best opposite price. Let us go back to the framework in which an agent is willing to buy a certain amount of a small tick asset. If the algorithm inserts an order inside the spread, there is no guarantee that between the moment the update message is sent by the exchange and the time the optimal distance is computed and the resulting order is sent, \textit{i.e.} which is commonly called the tick-to-trade latency, other agents have not quoted a new ask price inside the spread too. Hence, crossing this new best ask price would generate additional transaction costs since the limit order would instantly become a marketable order, incurring taker fees. This latency risk should be integrated in the cost function of the trading algorithm in order to penalize extreme aggressiveness.

To provide food for thought about modeling such a risk in our framework, let us denote by $\ell>0$ the tick-to-quote latency of the algorithm, \textit{i.e.} the time in seconds that separates the moment the message is sent by the venue, processed by the matching engine and the time the order resulting from the execution tactics is posted in the LOB. We suppose in the rest of the discussion that $\ell$ is negligible compared to the time horizon $T$, for example $\frac{\ell}{T}<10^{-3}$. This condition is not only necessary because the performance of high-frequency strategies vanishes with respect to latency, but is also convenient because we can assume
\begin{align}
    &\Delta p_\ell^a \ind L^{\delta,q},\\
    &\Delta p_\ell^a \ind \Delta p_T^a,
\end{align}
where $\square\ind\triangle$ stands for “$\square$ and $\triangle$ are independent random variables”.\\

We now denote by $\phi_\ell^z(x):=\mathbb{P}\left(\Delta p_\ell^a\leq x|Z=z\right)$ for a real number $x$ the cumulative distribution function of the best ask price variation over $[0,\ell]$ conditionally on a market state $z$. The latency-sensitive saved cost function $\mathcal{S_\ell}$ of the post and wait tactic is now written as
\begin{align} \label{eq:latency_sensitive_saved_cost_function}
    \mathcal{S}_\ell(T,\delta,q,z)&=\left(1-\phi_\ell^z\left(-(\psi_0+\delta)\right)\right)\mathcal{S}(T,\delta,q,z)\notag\\
    &\hspace{-1.75cm}-\phi_\ell^z(-(\psi_0+\delta))f^-\mathbb{E}_z\left[\Delta p_\ell^a|\Delta p_\ell^a\leq -(\psi_0+\delta)\right],
\end{align}
with $\mathcal{S}$ being given in Equation \eqref{eq:saved_cost_function}.

The last term represents the price discount of the marketable limit order over the immediate execution tactics due to a best ask price improvement. We see that the value of this new expected saved cost function is not necessarily smaller than the value of the latency-free one $\mathcal{S}$ but the maximum of the function is possibly attained at a different distance $\delta$.

\section{Discussion and conclusion}

In this work we introduced new microstructural features for fill probability computation, namely, the limit order flow imbalance, the aggressiveness index and the priority volume. We demonstrated their predictive power by exploring the smooth dependence of the fill and cancellation probability functions with respect to these features. We showed how neural networks with simple architectures can be used for the fill probability and clean-up cost computation using high-frequency data. A neural network was trained on a data base of real limit orders using a set of handcrafted interpretable features, and we analyzed the differences in the feature importance between CEXs cryptocurrency pairs and Euronext equities. We explained how taking into account both the priority volume and the distance of placement may be crucial in the case of small tick cryptocurrencies due to the sparse nature of their order book. Concerning the use of real limit orders over hypothetical orders, we discussed the main advantage in exploiting such informative data and strongly suggested to favor the real order flow other synthetic orders to eliminate the zero market impact assumption. By designing a cost function for an agent who aims at buying a quantity of the asset within a short time horizon, we demonstrated how to integrate such a model in a trading engine. A new backtest method was proposed, allowing to account for the market impact of limit orders using historical data only. We assessed the performance of the model and compared it to a toy model that involves a common form of the fill probability function. By computing the decisions the model would have made at the insertion of real limit orders, classification metrics can be used to study the relevancy of these execution models in the decision-making process. Finally, examples of optimal distance of placement were provided in the case of a small tick asset and numerical experiments suggest that the fee policy of the trading venue plays a decisive role concerning the aggressiveness of the order. Our findings suggest that in CEXs, provided that there is no latency, posting extremely aggressive limit orders is often optimal. The efficiency of such a radical tactic is explained by two main factors, the first one being the high fill probability of such an order and the second one being the fee policy that sets a significant difference between the maker fee and the taker fee. Hypothetically, inserting a bid limit order 1 tick below the best ask price when the spread is of the order of several hundreds of ticks would almost surely save 5 basis points of trading costs. In practice, such a tactic is prone to latency risk, and it can be integrated in the computation of the expected cost function.

In a multi-horizon framework, \textit{e.g.} execution algorithms with a trading horizon that can range from milliseconds to minutes depending on the market regime or operational constraints, the inference of the whole survival function is necessary. In this case, estimating more sophisticated models such as the ones from survival deep learning literature or the convolutional-transformer of \cite{arroyo2024deep} would be relevant. But more work needs to be done in order to propose an architecture that allows for fast computations in a live trading environment (computation time < 1 microsecond). Another extension of our approach would be to train a model for post-insertion evaluation. In a nutshell, the insertion of a limit order in the book bumps the order flow intensity and this excitation vanishes over time. This causes the fill probability of an order at its insertion to differ from the fill probability of a pending order with the exact same characteristics and the exact same set of features. The insertion of liquidity reveals information about the agent's intention and impacts the price, which is likely to move in the opposite direction. Such an extension can be carried out by simply adding pending limit orders in the training data and creating a new feature that characterizes the time elapsed since their insertion in the book. Last but not least, the study of the convexity of the saved cost function would highlight some regularity conditions that can be added as non-linear constraints in the loss functions of the neural network in order to improve the computation of the optimal distance.

\bibliography{main}
\bibliographystyle{apalike}

\section*{Appendix}
\label{sec:appendix}

\begin{figure}[!ht]
    \centering
    \subfloat[Contributions for passive orders $\delta > 0$, BTC-USD]{%
        \includegraphics[width=0.4\linewidth]{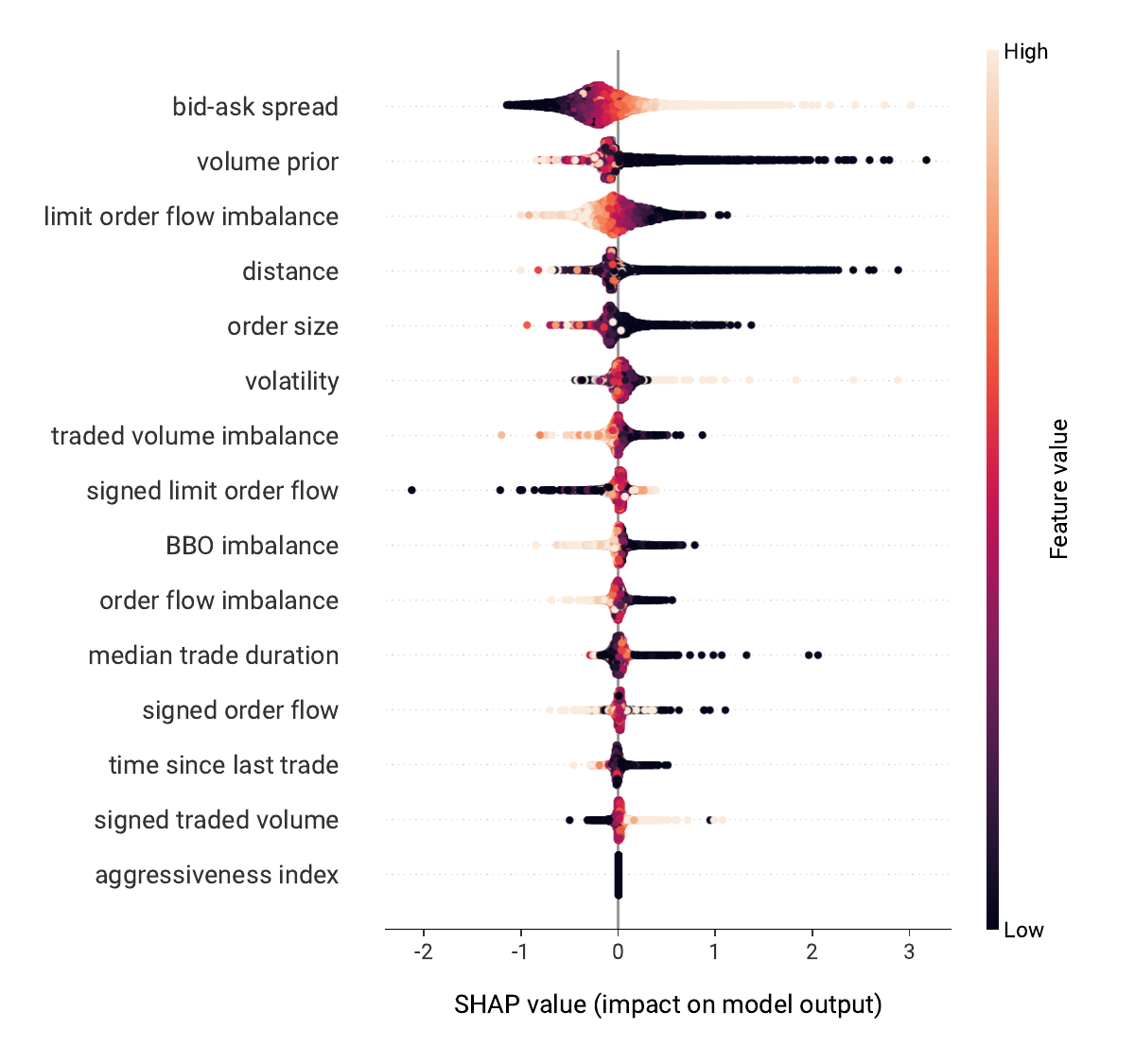}%
    }
    \subfloat[Contributions for passive orders $\delta > 0$, BNPP]{%
        \includegraphics[width=0.4\linewidth]{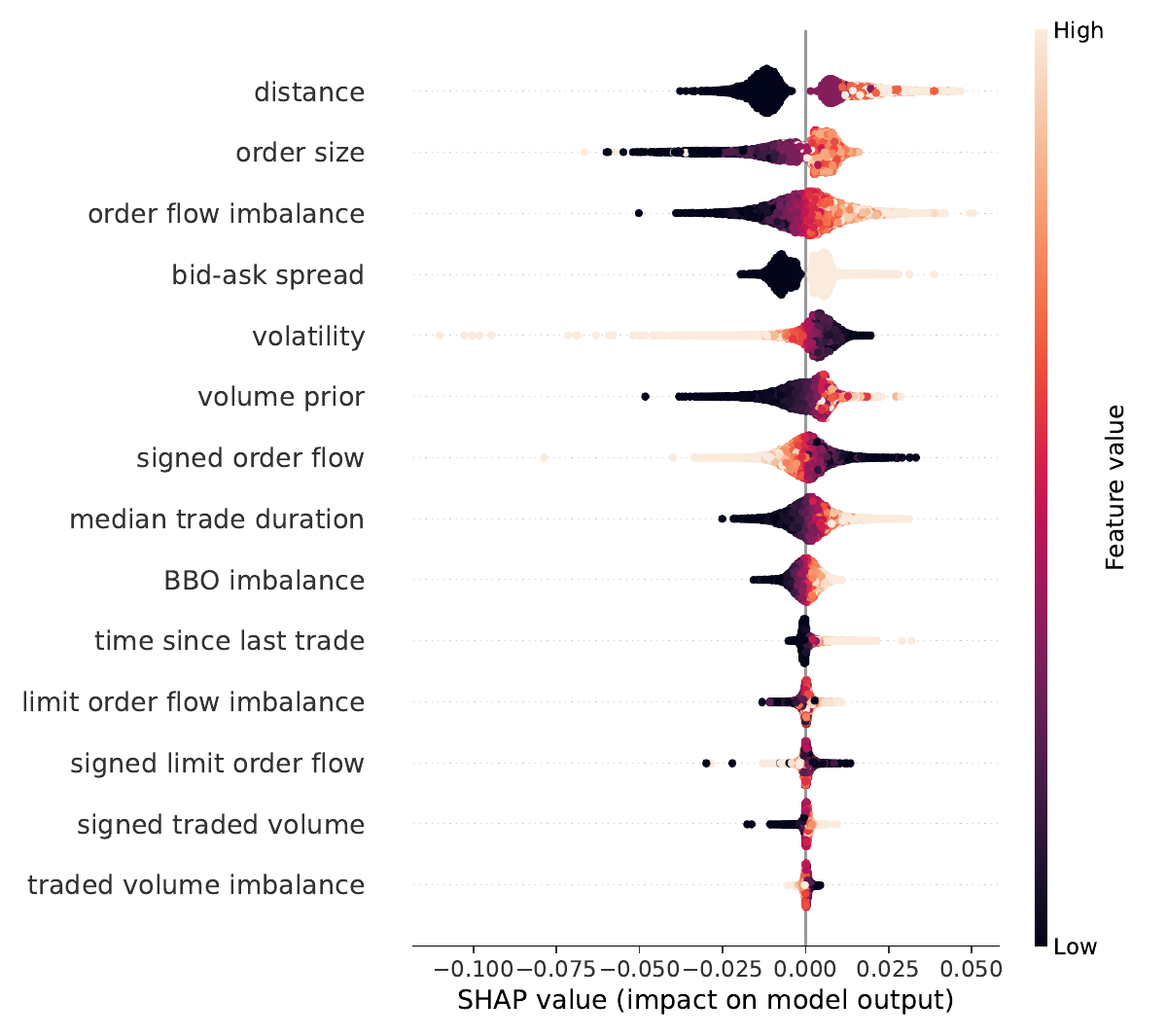}%
    }\\
    \subfloat[Contribution at best quote $\delta = 0$, BTC-USD]{%
        \includegraphics[width=0.4\linewidth]{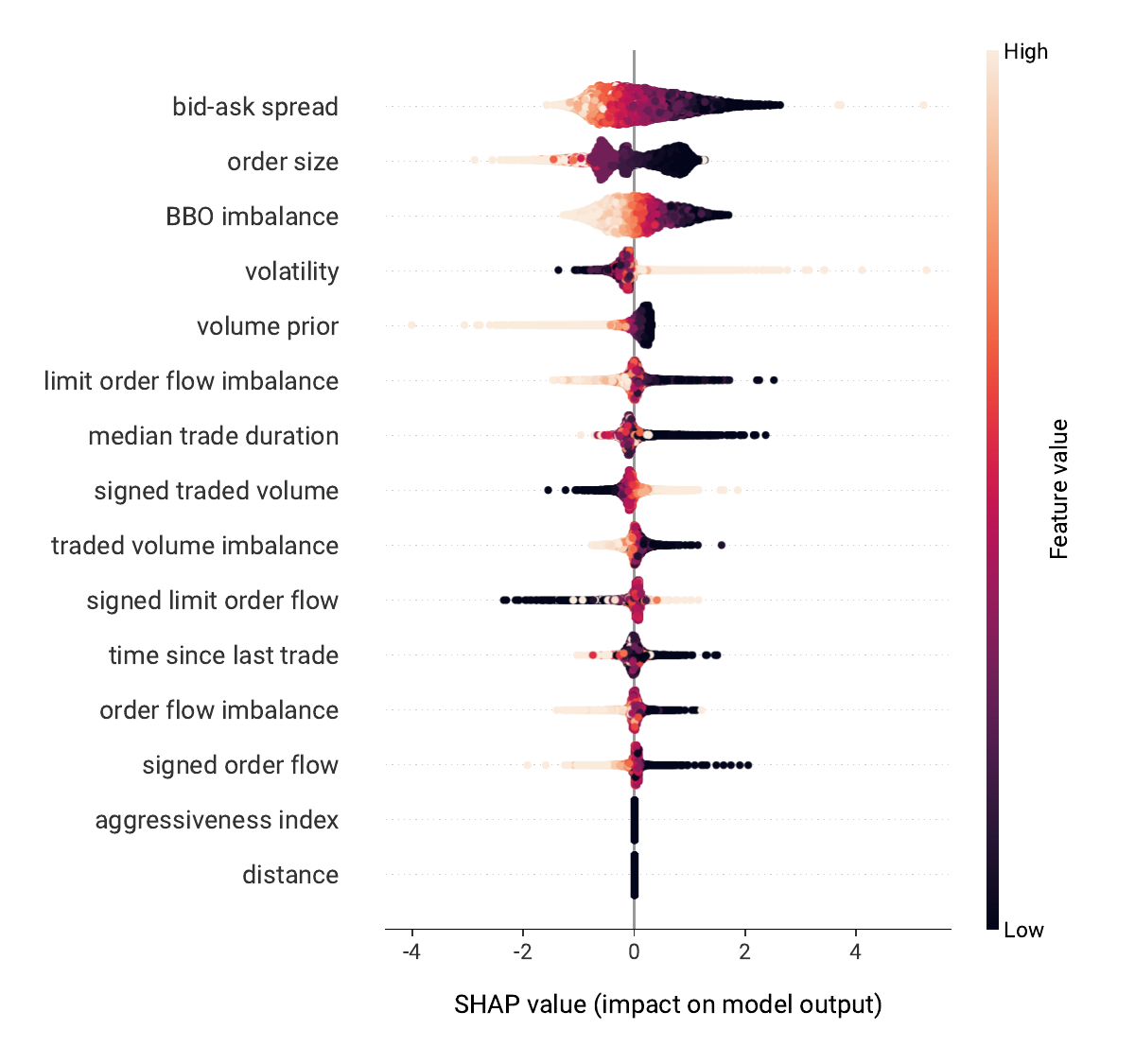}%
    }
    \subfloat[Contributions at best quote $\delta = 0$, BNPP]{%
        \includegraphics[width=0.4\linewidth]{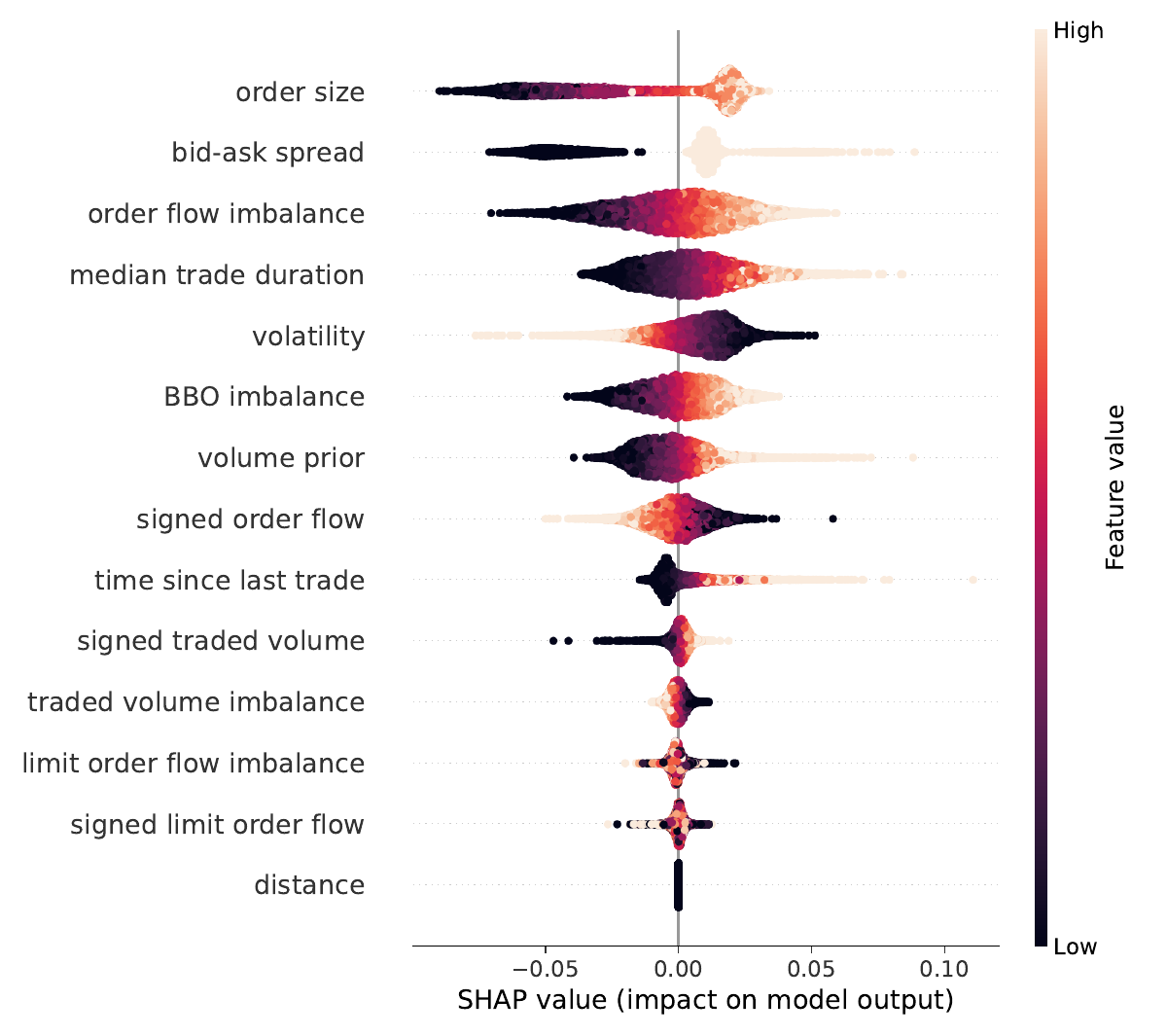}%
    }\\
    \subfloat[Contributions for aggressive orders $\delta < 0$, BTC-USD]{%
        \includegraphics[width=0.4\linewidth]{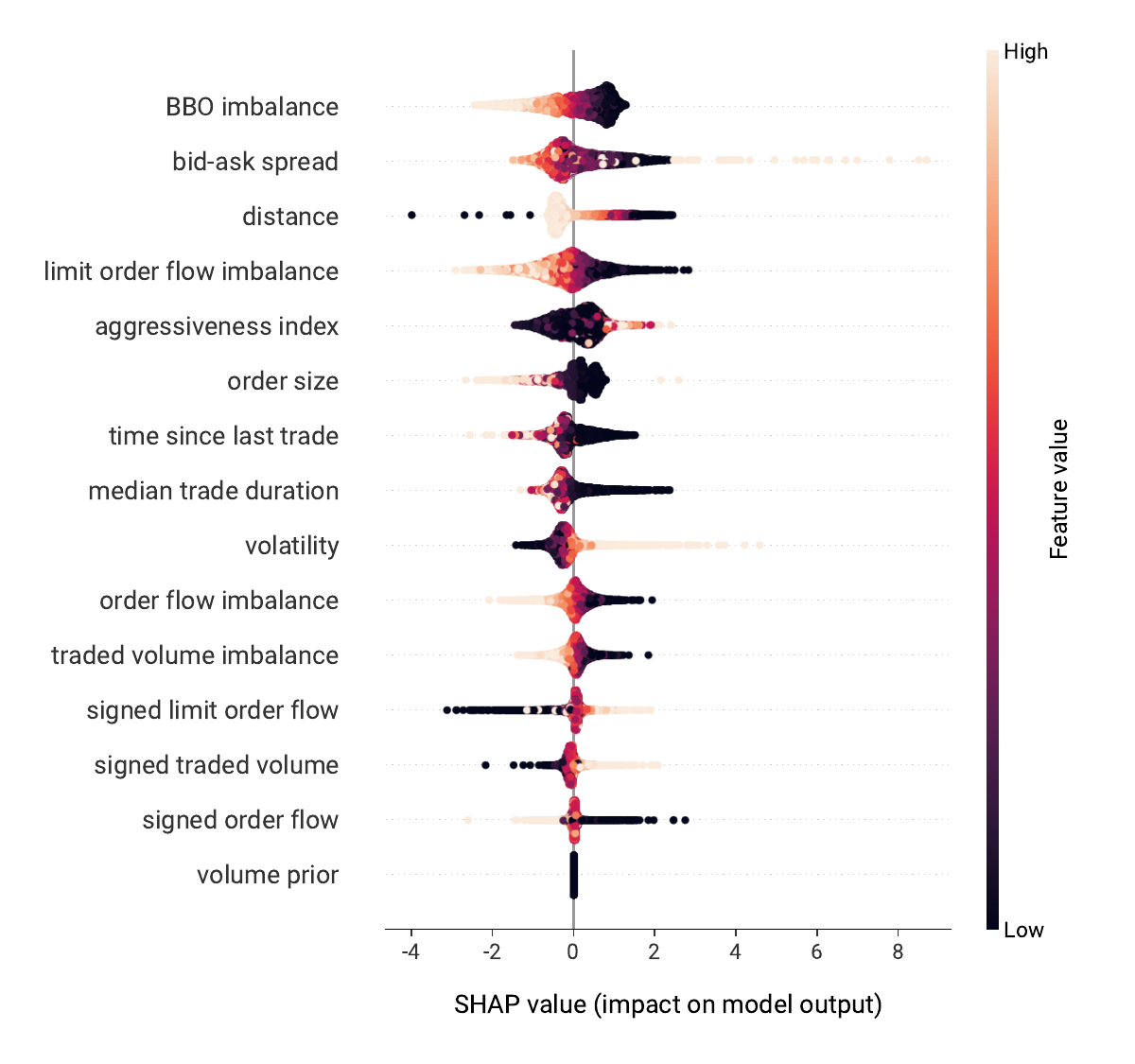}%
    }
    \subfloat[Contributions for aggressive orders $\delta < 0$, BNPP]{%
        \includegraphics[width=0.4\linewidth]{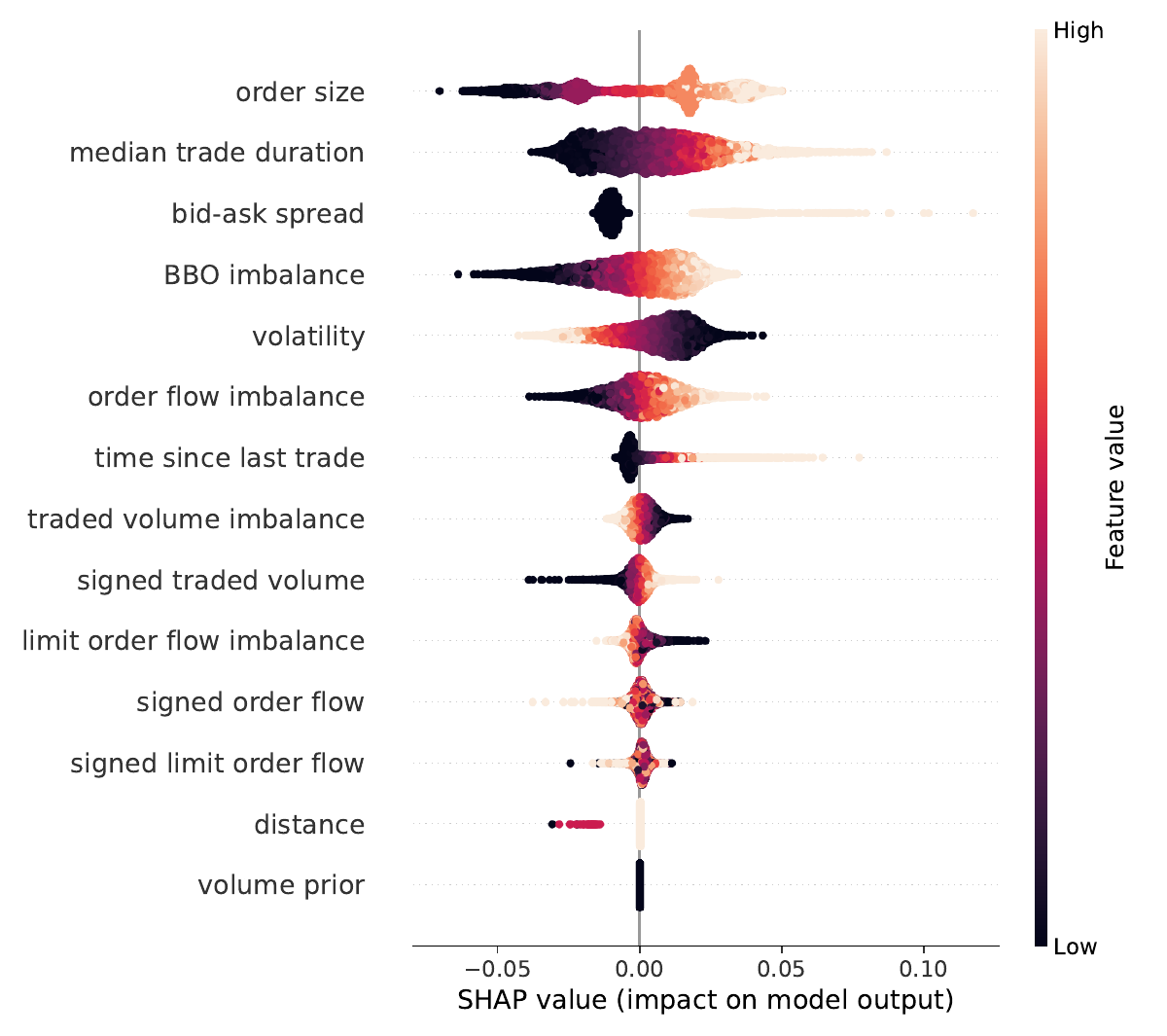}%
    }
    \caption{\textit{Feature importance} --- Market variables contributions to the saved cost function magnitude using Shapley values of 10,000 predictions, BTC-USD and BNPP, bid side.}
    \label{fig:shapley_saved_cost}
\end{figure}

\end{document}